\documentclass[aps,prd,superscriptaddress,showpacs,preprintnumbers]{revtex4}

\usepackage{epsfig}
\usepackage{graphicx,graphics}
\usepackage{amsmath}
\usepackage{amsfonts}
\usepackage{amssymb}

 \newcommand{\be}{\begin{eqnarray}}
 \newcommand{\ee}{\end{eqnarray}}
 \newcommand{\nee}{\nonumber\end{eqnarray}}
 \newcommand{\nn}{\nonumber\\}

  \newcommand{\bc}{\begin{center}}
 \newcommand{\ec}{\end{center}}
 \newcommand{\pup}{p^\uparrow}

\newcommand{\qup}{q^\uparrow}

\def\kt{k_\perp}
\def\bkt{\bf k_\perp}

\def\pp{p_\perp}

\def\pt{P_T}
\def\avk{\langle k_\perp ^2\rangle}
\def\avp{\langle p_\perp ^2\rangle}
\def\avPT{\langle P_T^2\rangle}

\def\S{_{_S}}
\def\T{_{_T}}
\def\C{_{_C}}

\def\BM{_{_{B\!M}}}

\def\xb{x_{_{\!B}}}

\def\s              {\sigma}
\def\g              {\gamma}

\begin{document}

\title{ Tests for and extraction of the Sivers,  Boer-Mulders and
transversity distributions in SIDIS reactions  }

\author{Ekaterina Christova}

\affiliation{ Institute for Nuclear Research and Nuclear Energy, Sofia 1784, Bulgaria}

\author{Elliot Leader}
\affiliation{  Imperial College London, London SW7 2AZ, UK}

\begin{abstract}
A major experimental program  is presently underway worldwide to determine the fundamental   non-perturbative  functions,
the Sivers, Boer-Mulders and transversity distributions, which are vital for an understanding of
the internal structure of the nucleon. However, at present, many simplifying assumptions are used in extracting   these
   functions from the data. We demonstrate that if the binning of the data in  $Q^2$ is small enough so that  $Q^2$-evolution can be neglected inside a bin, then one can  obtain stringent tests of these assumptions. Failure to satisfy these tests implies that the presently extracted non-perturbative  functions are unreliable.\newline
To this end we consider  the  measurement of the Sivers, Boer-Mulders and transversity \emph{difference}  asymmetries for
hadrons with opposite charges in SIDIS reactions with unpolarized
and transversely polarized deuteron and proton targets:  $l+N\to l'+h+X$, $h=\pi^\pm ,K^\pm ,h^\pm$. Utilizing
 only charge and isospin invariance, and applying the above mentioned simplifying assumptions we obtain several testable relations amongst the measured asymmetries. If these tests are satisfied then the measured
 asymmetries  determine two different combinations of the valence-quark transverse momentum dependent
 distributions,  which can be determined  separately without any contributions from the strange and other sea-quarks.
 \newline

\noindent
\end{abstract}


\maketitle

\section{Introduction}


The surprizingly large single-spin asymmetries found in
semi-inclusive reactions,  which were expected to be vanishingly
small on the basis of perturbative QCD, have been explained by
invoking relatively novel non-perturbative mechanisms: The Sivers and
Boer-Mulders parton distributions, and the transversity distribution
appear in conjunction  with the transverse momentum dependent
unpolarized and Collins fragmentation function. Knowledge of these
functions, important for understanding of the internal structure of
the nucleon, is obtained by  extracting them from experimental
data. At  present, this extraction has been relatively
simplistic, involving several key simplifying assumptions:

1) The analysis is in leading order in perturbative QCD

2) The transverse momentum dependent (TMD)  parton distributions (PDFs) and
fragmentation functions (FFs)
 which depend on parton intrinsic momentum $k_\perp $, are typically parametrized in a factorized form
 for the $\xb/z_h$ and $\kt$-dependences:
 \be
 \Delta f(\xb \, {\rm or}\, z_h, k_\perp^2)=\Delta f(\xb \,
{\rm or}\, z_h) \,e^{-k_\perp^2/\avk}\,\cdot
 \ee

 3) the $\xb (z_h)$-dependence is taken proportional to the collinear PDFs (FFs),

 4) the   Gaussian type $\kt $--dependence is assumed to be  flavour independent.

 The above  are the most commonly used simplified parametrizations. We shall refer to them as the {\it standard} parametrizations.
 There is presently  some discussion about the functional form of the TMD functions, and
 in a series of recent papers more general parametrizations have been suggested that allow
 flavour dependence of the partonic transverse momenta
 \cite{Bacchetta}, and models constructed for the
 perturbative transverse-momentum dependent  $Q^2$-evolution \cite{Q2-evol}, etc.
 Our aim is to suggest tests that would verify  to what
  extend  the  standard parametrizations provide  reliable information on the TMD distributions, and to show what information can be obtained if the tests are satisfied.

We show in the following that one can test these parametrizations
directly in semi-inclusive deep inelastic scattering (SIDIS),
 provided there is sufficient data to study, what we call, the "difference" cross section asymmetries, i.e. combinations of the type:
 \be
  A^{h-\bar h}\equiv \frac{\Delta \s^h-\Delta \s^{\bar h}}{\s^h-\s^{\bar h}}\label{A}
\ee
 where $\s^h$ and $\Delta\s^h=\s^{h\uparrow}-\s^{h\downarrow}$ are the unpolarized and polarized cross sections respectively.
 The arrows indicate the polarization of the target, and $h$ and $\bar h$ are hadrons with opposite charges.

 The  expressions for the difference asymmetries are  obtained using only charge (C) and isospin (SU(2)) invariance of the strong interactions.
 For kaons the additional assumption, made in all analysis,
that the unfavoured FFs of $d$ and $\bar d$-quarks into $K^+$ are the same, is used.
 There are two characteristic features of the difference cross sections:

 a) only the valence-quark parton densities and fragmentation function contribute and

  b)  the expressions factorize into a product of parton densities and fragmentation functions. The latter appear particularly important
  for the asymmetries, as common factors in the numerator and the denominator in (\ref{A}) cancel.

 The tests we consider become particularly simple for deutrons. For final hadrons
 $h=\pi^\pm, K^\pm $ and any unidentified
 charged hadrons $h^\pm$,
 the difference asymmetries always measure the sum of the valence quark contributions to each TMD function.
  The asymmetries on proton targets provide analogous tests,
  but  with different combinations of valence-quark TMD densities for pions and kaons.

 Presently, data on Sivers, Boer-Mulders (BM) and
  transversity asymmetries are  presented as functions of only one of the
kinematic variables $\xb,  z_h$, sometimes with, in addition, the
$Q^2 $ and  $P_T$ dependence, with integration over the measured
intervals of the other  variables.
 We obtain analytic expressions for the $\xb$ and $z_h$ dependence of the  asymmetries.
These expressions strongly simplify  if in the measured kinematic
intervals the binning in $Q^2$ is small enough to allow the neglect
 of the $Q^2$-dependence of the parton densities and  FFs.
 We shall  present both the general formulae and the formulae  where this simplification is valid.

  We shall demonstrate that measurements of the integrated  asymmetries  allow us to test the following assumptions currently used in addition to the standard simplifications
  in the extraction of the various non-perturbative functions:

 1. At present,  it is assumed that BM functions are proportional to the Sivers functions.
In Sect. \ref{Relations 1} we show that this can be tested since it
implies that the  BM and Sivers difference asymmetries have the same
the $\xb$-behaviour.

2. Since  the BM and transversity asymmetries  involve the same
Collins fragmentation functions, we show that there is a relation
between the $z_h$-behaviour of BM and Collins difference
asymmetries, which can be used as  a test of this basic assumption,
Sect. \ref{Relations 2}.

Previously\cite{we1,we2} we showed that the difference cross
sections for  SIDIS reactions with unpolarized and longitudinally
polarized targets
 with charged final hadrons,  in
the simple collinear picture involve only the valence quark parton
densities and fragmentation functions. Later we showed that
 this holds for  unpolarized SIDIS in the non-collinear approach involving TMD parton densities and
fragmentation functions as well \cite{we2}.

 Here we extend our studies to the  Sivers, Boer-Mulders and
transversity asymmetries
 for the difference cross sections in SIDIS of unpolarized leptons on a transversely polarized \emph{deuteron} and \emph{proton} targets.
These asymmetries  allow   tests of the  commonly  used standard simplifications.
If the analysis survive the tests, then measurements on deutrons provide direct
information on the sum of the valence-quark BM, Sivers and
transversity distributions, and  measurements on protons provide information about another combination of the
valence-quark TMD densities: on $(e_u^2\,\Delta f_{u_V}-e_d^2\, \Delta f_{d_V})$ with final pions, and on
$\Delta f_{u_V}$ with final kaons.
Measurements on both protons and deutrons
will determine the separate valence-quark densities
without any contributions from the sea-quark densities. A first attempt to use this approach  on the single-spin  difference asymmetry for charged pions
  was recently presented by the HERMES collaboration~\cite{HERMES_2009} in an effort
   to isolate the valence-quark Sivers functions.

\section{Sivers distribution functions}

The Sivers distribution function $\Delta^N \! f_{q/\pup}(\xb, \kt)$
appears in the expression for
  the number density of unpolarized quarks $q$ with intrinsic transverse momentum
 ${\bf k}_\perp$ in a
 transversely polarized proton $\pup$ with 3-momentum $\bf P$ and transverse spin $\bf S_T$ \cite{Sivers}:
\be
f_{q/\pup}(\xb, {\bf \kt}) = f_{q/p}(\xb, \kt) +\frac{1}{2}\,\Delta^N \!
f_{q/\pup}(\xb, \kt) \,\,
 {\bf S_T}\cdot ({\bf \hat P}\times {\bf \hat k}_\perp)
\ee
where $f_{q/p}(\xb, \kt)$ are the unpolarized $\xb$ and
$k_\perp$-dependent parton densities and the triple product induces
a definite azimuthal $\sin (\phi_h-\phi_ S)$-dependence,
$\phi_h$ and $\phi_S$ are the azimuthal angles of
the final hadron and the spin of the target,
  $x_B=Q^2/2(P.q)$ is the usual Bjorken variable.

 Often the notation $f_{1T}^{\perp q}(\xb, \kt) $ \cite{MuldersTangerman} and $\Delta f_{q/S_T}(\xb, \kt)$ \cite{Anselmino_06} is also used:
 \be
\Delta^N \! f_{q/\pup}(\xb, \kt)=\Delta f_{q/S_T}(\xb, \kt)=
 - \frac {2\kt}{m_p} \, f_{1T}^{\perp q} (\xb, \kt)  \label{siv}
\ee
 where $m_p$ is the proton mass.
\section{Sivers  difference asymmetries }

To access the Sivers TMDs one considers the $\sin(\phi_h-\phi_S)$
azimuthal moment of the  transverse single-spin
$A_{UT}^{\sin(\phi_h-\phi_S)}$ asymmetry in SIDIS \cite{general}:
\be
A_{UT}^{\sin(\phi_h-\phi_S),h}&=&\frac{1}{S_T}\, \frac{\int
d\phi_h d\phi_S \left[d^6 \s^\uparrow-d^6\s^\downarrow
\right]^{h}\,\sin(\phi_h-\phi_S)} {\int d\phi_h d\phi_S \left[d^6
\s^\uparrow+d^6\s^\downarrow \right]^{h}}\nn
&=&\frac{1}{2}\,\frac{\,\frac{A(y)}{Q^4}\,F_{UT}^{\sin(\phi_h-\phi_S),h}}
{\,\frac{A(y)}{Q^4}\,F_{UU}^h},\qquad A(y)=[1+(1-y)^2]\,\cdot
\ee
 Here  $d^6\s^{\uparrow,\downarrow}$ stands for the differential
cross section of SIDIS with
 an unpolarized lepton beam on a  transversely polarized target
 in the kinematic region  $P_T \simeq k_\perp \ll Q$  at
  order $(k_\perp /Q)$, the arrows ($\uparrow,\downarrow$)
stand for the transverse target polarization:
 \be
\left[d^6\s^{\uparrow,\downarrow}\right]^h\equiv\frac{d\s^{\ell
\,p^{\uparrow,\downarrow} \to \ell^\prime h X}}
{d\xb\,dQ^2\,dz_h\,dP_T^2\,d\phi_h\,d\phi_S},
\ee
  $P_T$ is the transverse momentum of the final hadron  in the
$\g^*-p$\, c.m. frame,
 and $ z_h$, $Q^2$ and $y$ are the usual measurable SIDIS quantities:
\be
 \quad z_h=\frac {(P.P_h)}{(P.q)},\quad Q^2=-q^2, \quad q=l-l',
\quad
 y=\frac{(P.q)}{(P.l)}
 \ee
 with $l$ and $l'$, $P$ and $P_h$  the 4-momenta of the initial
and final leptons,  and initial and final hadrons.  Note that
 \be
 \quad Q^2=2ME \xb y
\ee
where $M$ is the target mass (in this paper the deuteron mass) and $E$ the lepton laboratory energy.
Throughout the
paper we
 follow the notation and kinematics of ref. \cite{general}.

The Sivers asymmetry for the difference cross section we define
analogously:
\be
 A_{UT}^{Siv,{h-\bar h}}(\xb , Q^2,z_h,
P_T^2)&=&\frac{1}{S_T}\, \frac{\int d\phi_h d\phi_S \left[d^6
\s^\uparrow-d^6\s^\downarrow \right]^{ h-\bar
h}\,\sin(\phi_h-\phi_S)} {\int d\phi_h d\phi_S \left[d^6
\s^\uparrow+d^6\s^\downarrow \right]^{h-\bar h}}\nn
&=&\frac{1}{2}\,\frac{\frac{A(y)}{Q^4}\,F_{UT}^{\sin(\phi_h-\phi_S),h-\bar
h}} {\frac{A(y)}{Q^4}\,F_{UU}^{h-\bar h}}
\ee

The structure functions $F_{UT}^{\sin(\phi_h-\phi_S),h-\bar h}$
involve convolutions of
 the Sivers and unpolarized TMD fragmentation functions,while the
 $F_{UU}^{h-\bar h}$ are convolutions of the unpolarized TMD parton densities and fragmentation functions.
   We consider the difference asymmetries on  deuteron and proton targets separately.


\subsection{The standard parametrizations for deutrons}

In general, in the difference cross sections only the valence-quark
TMD functions enter. When a deuteron target is used,  a further
simplification occurs -- independently of the final hadrons,  only
one parton density enters -- the sum of the valence quarks:
\be
\Delta f_{Q_V/S_T} (\xb,\kt)\equiv \Delta f_{u_V/S_T} (\xb,\kt)+\Delta f_{d_V/S_T} (\xb,\kt)
\ee

In extracting the Sivers distributions from experiment they are
assumed proportional to the unpolarized densities
$f_{q/p}(\xb,k_\perp)$ with  additional $\xb$- and
$k_\perp$-dependencies in a factorized form \cite{general}.
Following this convention, we adopt the following  parametrization
for valence-quark Sivers function:
\be
 \label{MS} \Delta f_{Q_V/S_T} (\xb,\kt)=\Delta
f_{Q_V/S_T} (\xb)\,\sqrt{2e}\, \frac{\kt}{M\S } \;
\frac{e^{-\kt^2/\avk\S }}{\pi\avk} \label{Siv-dist}
 \ee
 where
  \be
   \label{curlyN}
   \Delta f_{Q_V/S_T} (\xb,Q^2)=2\,{\cal N}^{Q_V}_{Siv}(\xb)\;Q_V(\xb,Q^2)   \ee
   and $Q_V$  is the sum of the  collinear valence PDFs:
\be
   Q_V=u_V+d_V\,\cdot \ee
 $ M\S $ is an unknown,  fitting parameter (sometimes denoted by $M_1 $ \cite{Anselmino_08}),   ${\cal N}^{Q_V}_{Siv}(\xb)$ is an unknown function
 and
 \be
  \label{kperpSiv}
 \quad \avk\S  = \frac{\avk  \, M^2\S }{\avk  +  M^2\S }\,\cdot
   \ee

The parameter $\avk $ enters  the \emph{unpolarized}
parton densities $f_{Q_V/p}(\xb,k_\perp , Q^2)$, which are
conventionally
 expressed in terms of the collinear parton densities times a
   Gaussian-type  transverse momentum dependent function:
\be
f_{Q_V/p}(\xb,k_\perp ,
Q^2)&=&Q_V(\xb,Q^2)\,\frac{e^{-k_\perp^2/<k_\perp^2>}}{\pi<k_\perp^2>}\,\cdot\label{fq}
\ee
 Here  $\avk $ is
assumed known  from a study of multiplicities in unpolarized SIDIS \cite{multpls}.
On the other hand  ${\cal N}^{Q_V}_{Siv}(\xb)$ and $M\S $ in
Eqs.~(\ref{MS}) and (\ref{curlyN})  are unknowns.

Analogously, the unpolarized FFs
 are conventionally factorized into a product of the collinear FFs times
 a   Gaussian-type  transverse momentum dependent function:
 \be
D_{q_V}^h(z_h,p_\perp ,Q^2) &=&D_{q_V}^h(z_h,Q^2)
\,\frac{e^{-p_\perp^2/\avp }}{\pi\avp },\quad q=u,d\label{Dq}
\ee
 where
$D_{q_V}^h(z_h)$ are the  collinear valence  FFs.
 Multiplicities provide information about the parameter $\avp $.


\subsection{The structure functions for deuterons}

With the above parametrisations we obtain the  expressions for
$F_{UT}^{\sin(\phi_h-\phi_S),h-\bar h}$. They differ
 for the different final hadrons \emph{only} by the known factorized  collinear FFs:
\be
F_{UT}^{\sin (\phi_h -\phi_S),\pi^+-\pi^-} & = & {\cal K} \S
(z_h,P_T^2)  \,
 \Delta f_{Q_V/S_T}(\xb,Q^2)\,(e_u^2-e_d^2)\,
  D^{\pi^+}_{u_V}(z_h)\label{FUTSiv-pions}\\
F_{UT}^{\sin (\phi_h -\phi_S),K^+-K^-} & = & {\cal K} \S (z_h,
P_T^2)  \,
 \Delta f_{Q_V/S_T}(\xb,Q^2)\,e_u^2\,
 D^{K^+}_{u_V}(z_h)\label{FUTSiv-kaons}\\
F_{UT}^{\sin (\phi_h -\phi_S),h^+-h^-} & = & {\cal K} \S (z_h,
P_T^2)  \,
\Delta f_{Q_V/S_T}(\xb,Q^2)\,[\,e_u^2\,
 D^{h^+}_{u_V}(z_h)+e_d^2\,  D^{h^+}_{d_V}(z_h)]
\ee
where ${\cal K}\S(z_h,P_T^2)$ is a common flavour independent
factor:
  \be
   {\cal
K}\S(z_h,P_T^2)&=&\frac{\sqrt{2e}}{2}\,A_{Siv}\,  P_T \,
\frac{e^{-P_T^2/\avPT_{Siv} }}{\pi\avPT_{Siv} ^2 }\,z_h ,\qquad A_{Siv}=\frac{
\avk\S  ^2 }{M\S \,\avk},\label{ASiv}
 \ee
  \be
   \avPT\S
=\avp +z_h^2\avk\S .\label{PperpSiv}
 \ee

We recall the expressions for $F_{UU}^{h-\bar h}$ that enter as
normalizing factors in the considered asymmetry \cite{we2}:
 \be
 F_{UU}^{\pi^+ - \pi^-} &=&Q_V(\xb
)\,(e_u^2-e_d^2)\,D_{u_V}^{\pi^+}(z_h)\,\frac{e^{-P_T^2/\langle
P_T^2\rangle }}{\pi\,\langle P_T^2\rangle }\label{FUU-pions}\\
 F_{UU}^{K^+ - K^-} &=&Q_V(\xb )\,e_u^2\,D_{u_V}^{K^+}(z_h)\,\frac{e^{-P_T^2/\langle
P_T^2\rangle}}{\pi\,\langle P_T^2\rangle}\label{FUU-kaons}\\
 F_{UU}^{h^+ - h^-}
&=&Q_V(\xb )\left[\,e_u^2\,D_{u_V}^{h^+}(z_h)+
e_d^2\,D_{d_V}^{h^+}(z_h)\right] \,\frac{e^{-P_T^2/\avPT}}{\pi\,\langle P_T^2\rangle}
 \ee
 where
 \be
\avPT =\avp +z_h^2\avk  \,\cdot \label{PT}
\ee

Thus, the Sivers difference asymmetries  $A_{UT}^{Siv,{ h-\bar h}}$,
$h=\pi^+,K^+,h^+$, independently of the final hadrons,
 determine  the same valence-quark Sivers function $\Delta f_{Q_V/S_T}(\xb)$ without any
contributions from the sea-quark Sivers densities. The only unknowns
are  $M\S $ (involved in the definition of $\avk\S$)
 and ${\cal N}^{Q_V}_{Siv}(\xb )$.

The data on Sivers asymmetries are usually  presented as function of
only one of the kinematic variables $(\xb,z_h,P_T)$, integrated over
the measured intervals of the other  variables. In the following we
consider the $\xb$ and $z_h$-dependent Sivers asymmetries. The
expressions  simplify dramatically if we assume that the binning of
the data in $Q^2$ is fine enough so that we can neglect
 the $Q^2$-dependence of the \emph{collinear functions} inside each bin.

\subsection{ The integrated Sivers asymmetries on deutrons}

We perform the $P_T$ integration analytically.
 The general expression is presented in the
Appendix.
Integrating further we obtain the $\xb$- and $z_h$-dependent Sivers asymmetries.\\

\subsubsection{The $z_h$-dependent Sivers  asymmetries}

  For the
$z_h$-dependence of Sivers  difference asymmetry for $h=\pi^+,K^+,h^+$ we obtain:
\be
A_{UT}^{Siv, h-\bar h}(z_h ) &=&\frac{1}{2}\,\frac{\int dQ^2\,d\xb
\,dP_T^2\, \frac{A(y)}{Q^4}\,F_{UT}^{\sin(\phi_h-\phi_S),{ h-\bar
h}}(\xb ,Q^2,z_h,P_T^2)} {\int dQ^2\,d\xb
\,dP_T^2\,\frac{A(y)}{Q^4}\,F_{UU}^{ h-\bar h}(\xb
,Q^2,z_h,P_T^2)}\nn
 &=& {\cal
B}_{Siv}^{h}(z_h)\,\frac{z_h}{\sqrt{\avp+z_h^2\avk\S }},\label{A_S_general}
\ee
 Here $ {\cal B}_{Siv}^{h}(z_h)$  for
$h=\pi^+, K^+$ are:
 \be
 {\cal B}_{Siv}^{h}(z_h)=\frac{\sqrt{e\pi}}{4\sqrt{2}}\, A_{Siv}\,\frac{\int d\xb\,\int
dQ^2\, \frac{1+(1-y)^2}{Q^4}\,\Delta
f_{Q_V/S_T}(\xb,Q^2)\,D_{u_V}^{h}(z_h,Q^2)}
 {\int d\xb\,\int dQ^2\,\frac{1+(1-y)^2}{Q^4}\,Q_V(\xb,Q^2)\,D_{u_V}^{h}(z_h,Q^2)}.
 \ee
For unidentified charged hadrons $h^\pm$, $ {\cal
B}_{Siv}^{h^+}(z_h)$ is obtained
 via the following replacement:
\be
 D_{u_V}^{h^+}(z_h,Q^2)\,\to \, e_u^2 D_{u_V}^{h^+}(z_h,Q^2) +
e_d^2 D_{d_V}^{h^+}(z_h,Q^2)\label{replacement}
\ee

For bins corresponding to a reasonably small interval $\Delta Q^2$ in $Q^2$, we replace the integral
over $Q^2$ by $\Delta Q^2 $ times the $Q^2$-dependent functions evaluated at the mean value $\bar {Q}^2$ for the bin. Then the FFs cancel and (\ref{A_S_general})
becomes particularly simple:
\be
 A_{UT}^{Siv, h-\bar h}(z_h,\bar
{Q^2} )=
 {\bar B}_{Siv}(\bar {Q^2})\,\frac{z_h}{\sqrt{\avp +z_h^2 \avk\S }}.\label{ASivzd}
\ee
 where $ {\bar B}_{Siv}$ is independent of $z_h$ and $h$:
\be
 {\bar B}_{Siv}(\bar
{Q^2})=\frac{\sqrt{e\pi}}{4\sqrt{2}}\, A_{Siv}\,\frac{\int d\xb\,\,[1+(1-\bar{y})^2] \, \Delta f_{Q_V/S_T}(\xb,\bar{Q^2})}
{\int d\xb\,\,[1+(1-\bar{y})^2]\,Q_V(\xb,\bar
{Q^2})}.\quad \forall\, h\label{BSivzd}
\ee
and
\be
 \bar{y} = \frac{\bar{Q}^2}{2MEx_B} \label{y}
\ee

 Thus, (\ref{ASivzd})
  is a remarkably strong prediction both, for the explicit $z_h$-behaviour and for the independence on $h$.
 It  provides a stringent test of the assumptions made in the extraction of the Sivers function from experiment.
Failure to satisfy the test implies that the present  information on the Sivers function is unreliable.
  If the test is satisfied one obtains
 straightforward information about $\avk\S $.

 \subsubsection{The $\xb$-dependent Sivers  asymmetries}

   The  $\xb$-dependent Sivers difference asymmetry  is directly proportional to
  ${\cal N}^{q_V}_{Siv}(\xb )$:
\be
A_{UT}^{Siv,{ h-\bar h}}(\xb )&=&\frac{1}{2}\,\frac{\int
dQ^2\,dz_h \,dP_T^2\,
\frac{A(y)}{Q^4}\,F_{UT}^{\sin(\phi_h-\phi_S),{ h-\bar h}}(\xb
,Q^2,z_h,P_T^2)} {\int dQ^2\,dz_h
\,dP_T^2\,\frac{A(y)}{Q^4}\,F_{UU}^{ h-\bar h}(\xb
,Q^2,z_h,P_T^2)}\nn &=& {\cal C}_{Siv}^{h}(\xb)\,{\cal
N}^{Q_V}_{Siv}(\xb )\,,
 \ee
  where ${\cal C}_{Siv} ^{h}(\xb)$, $h=\pi^+, K^+$ is given by:
 \be
 \hspace*{-1cm} {\cal C}_{Siv} ^{h}(\xb)=\frac{\sqrt{e\pi}}{2\sqrt{2}}\, A_{Siv}\frac{\int dQ^2\,\frac{1+(1-y)^2}{Q^4}\,Q_V(\xb,Q^2)\,\int
dz_h \,z_h \left[ D_{u_V}^{h}(z_h, Q^2)\right]/\sqrt{\avPT\S  }} {\int
dQ^2\,\frac{1+(1-y)^2}{Q^4}\,Q_V(\xb,Q^2)\,\int dz_h
\,D_{u_V}^{h}(z_h, Q^2)} \,\cdot
\ee
For unidentified charged hadrons
$h^\pm$, $ {\cal C}_{Siv}^{h^+}(z_h)$ is obtained via
replacement (\ref{replacement}).

For small enough bins in $Q^2$, as discussed in Section~(2.3.1),   the asymmetry becomes:
 \be
A_{UT}^{Siv,{ h-\bar h}}(\xb,\bar {Q^2} )= \bar C_{Siv}^{h}
(\bar{Q^2})\,{\cal N}^{Q_V}_{Siv}(\xb )\,,\label{xb_S_approx}
 \ee
where $\bar C_{Siv}^{h}$ are independent of $\xb$ and known from
multiplicities.
\be
 \bar C_{Siv}^{h}(\bar{Q^2})=\frac{\sqrt{e\pi}}{2\sqrt{2}}\, A_{Siv}\,\frac{\int dz_h \,z_h
\left[ D_{u_V}^{h}(z_h,\bar {Q^2})\right]/\sqrt{\avPT\S  }} {\int dz_h \,D_{u_V}^{h}(z_h,\bar {Q^2})},\quad
h=\pi^+, K^+\label{CSivxd}
 \ee
and
\be
\bar C_{Siv}^{h^+}(\bar{Q^2}) &=&\frac{\sqrt{e\pi}}{2\sqrt{2}}\, A_{Siv}\,\frac{\int dz_h\,z_h
\left[e_u^2 D^{h^+}_{u_V}+e_d^2 D^{h^+}_{d_V}\right]/\sqrt{\avPT\S  }} {\int dz_h \left[e_u^2 D^{h^+}_{u_V}+e_d^2
D^{h^+}_{d_V}\right]}\,\cdot
\ee
where $\avPT\S $ is given in Eq.~(\ref{PperpSiv}).

It follows from  Eq.~(\ref{xb_S_approx}) that the ratio for
different hadrons  should be independent of $\xb$ -- a very strong
test of
 the commonly used assumptions for  extracting the TMD functions.\newline
In summary  the tests above are important for assessing the reliability of the extraction of the Sivers function and are a strong motivation
 for measurements using reasonably small bins in $Q^2$ where the slow $Q^2$-dependence of the PDFs and FFs can be neglected.
 In addition, if the tests are satisfied one obtains  direct information on ${\cal N}^{Q_V}_{Siv}(\xb )$.


\subsection{Sivers difference asymmetries on protons }

 Crucial for the  tests on a deuteron target was that the PDFs and FFs factorize, i.e. there is no sum over quark flavours.
For a deuteron target this was a general feature valid for all final hadrons. For a
proton target this holds only for final pions and kaons $(\pi^+-\pi^-)$ and $(K^+-K^-)$, and does not hold for unidentified final hadrons.

Hence we consider Sivers difference asymmetries on protons only for $\pi^\pm$ and $K^\pm$.
The corresponding expressions are obtained from the expressions for deuteron target with the following replacements:

$\bullet$ for $\underline{\underline{\pi^+-\pi^-}}$

In the numerators, determined by  $F_{UT}^{\sin (\phi_h-\phi_S),\pi^+-\pi^-}$ (eq. (\ref{FUTSiv-pions})):
\be
(e_u^2-e_d^2)\,\Delta f_{Q_V/S_T}(\xb,Q^2)\, \rightarrow \,\Delta f_{Siv}^V(\xb ,Q^2)
\ee
 where
 \be
\Delta f_{Siv}^V(\xb ,Q^2)\equiv 2\left[\,e_u^2\,{\cal
N}_{Siv}^{u_V}(\xb)\,u_V(\xb,Q^2)-e_d^2\,{\cal
N}_{Siv}^{d_V}(\xb)\,d_V(\xb,Q^2)\right]
\ee
and in the denominator,
determined by $F_{UU}^{\pi^+-\pi^-}$ (eq, (\ref{FUU-pions})):
 \be
(e_u^2-e_d^2)\,Q_V(\xb,Q^2)\, \rightarrow \,f^V(\xb
,Q^2)\label{pions} \ee
 where
 \be
 f^V(\xb ,Q^2)\equiv \,e_u^2\,u_V(\xb,Q^2)-e_d^2\,d_V(\xb,Q^2)\cdot\label{fV}
\ee

$\bullet$ for $\underline{\underline{K^+-K^-}}$

In the numerators, determined by $F_{UT}^{\sin (\phi_h-\phi_S),K^+-K^-}$, eq. (\ref{FUTSiv-kaons}):
\be
\Delta f_{Q_V/S_T}(\xb,Q^2)\to \Delta f_{u_V/S_T}(\xb,Q^2)
\ee
where
\be
\Delta f_{u_V/S_T}(\xb,Q^2)=2\,{\cal N}_{Siv}^{u_V}(\xb)\,u_V(\xb,Q^2),
\ee
and in the denominator, determined by $F_{UU}^{K^+-K^-}$ (eq, (\ref{FUU-kaons})):
\be
Q_V(\xb,Q^2)\, \rightarrow \,u_V(\xb,Q^2)\cdot\label{K}
\ee

We shall not present all results, but only those  for small enough ranges of $Q^2$,
 where the $Q^2$-evolution of $q_V$ and
$D_{u_V}^{h}$ can be neglected.\\

{\bf 1a)} The $z_h$-dependent asymmetries on protons  have
 the same $z_h$-behaviour for $\pi^+-\pi^-$ and $K^+-K^-$. Moreover, this behaviour is the same as for deuteron targets
 and reflects the chosen Gaussian dependence on $\kt$ and $p_\perp$.
 Instead of (\ref{ASivzd}) and (\ref{BSivzd}) for deuterons, the analogous expressions for protons  are:
\be
A_{UT,p}^{Siv, \pi^+-\pi^-}(z_h,\bar {Q^2} )&=&
 \bar B_{Siv}^{\pi}(\bar {Q^2})\,\frac{z_h}{\sqrt{\avp +z_h^2 \avk\S}}\nn
 A_{UT,p}^{Siv, K^+-K^-}(z_h,\bar {Q^2} )&=&
  \bar B_{Siv}^{K}(\bar {Q^2})\,\frac{z_h}{\sqrt{\avp +z_h^2 \avk\S}}.\label{S-zh-approx_p}
\ee
They differ only in the constant common factors:
\be
 \bar B_{Siv}^{\pi}(\bar {Q^2})&=&\frac{\sqrt{e\pi}}{2\sqrt{2}}\, A\S\,\frac{\int d\xb\,
(1+(1-\bar y)^2)\,\Delta f_{Siv}^V(\xb,\bar{Q^2})} {\int
d\xb\,(1+(1-\bar y)^2)\,f^V(\xb,\bar {Q^2})}\nn
  \bar B_{Siv}^{K}(\bar {Q^2})&=&\frac{\sqrt{e\pi}}{\sqrt{2}}\, A\S\,\frac{\int d\xb\,
(1+(1-\bar y)^2) \,{\cal N}_{Siv}^{u_V}(\xb)u_V(\xb,\bar{Q^2})} {\int
d\xb\,(1+(1-\bar y)^2)\,u_V(\xb,\bar {Q^2})}
\ee

{\bf 1b)} The $\xb$-dependent Sivers asymmetries on protons are analogous to those on deuterons,
but with  different combinations of valence-quark densities for pions and kaons.
We have:
\be
A_{UT,p}^{Siv,{ \pi^+-\pi^-}}(\xb,\bar {Q^2} )&=&
  \bar C_{Siv}^{\pi} (\bar{Q^2})\,\frac{\Delta f_{Siv}^V(\xb,\bar{Q^2})}{2\,f^V(\xb,\bar{Q^2})}\nn
A_{UT,d}^{Siv,{K^+-K^-}}(\xb,\bar {Q^2} )&=&  \bar C_{Siv}^{K} (\bar{Q^2})\,{\cal N}^{u_V}_{Siv}(\xb )\label{xb_S_approx_p}
\ee
where $f^V(\xb)$ is known from DIS, eq. (\ref{fV}). They provide information about $\Delta f_{Siv}^V$ and ${\cal N}_{Siv}^{u_V}$.
Note that the coefficients $\bar C_{Siv}^{h} $ are the same as for deuterons, eq.(\ref{CSivxd}).
This allows us to form different ratios, for example:
\be
\left[\frac{A_{UT,d}}{A_{UT,p}}\right]^{\pi^+-\pi^-}&=&\frac{{\cal N}_{Siv}^{Q_V}(\xb)}{\Delta f_{Siv}^V(\xb,\bar{Q^2})}\, 2 f^V(\xb,\bar Q^2)\nn
\left[\frac{A_{UT,d}}{A_{UT,p}}\right]^{K^+-K^-}&=&\frac{{\cal N}_{Siv}^{q_V}(\xb)}{{\cal N}_{Siv}^{u_V}(\xb)}
\ee
in which the r.h. sides are independent of the transversity parameters $\avk\S,\avk,\avp$.

It should be remembered however, that in contrast to deuteron targets, these results do not hold for unidentified charged hadrons.



\section{Boer-Mulders distributions}
The
extraction of the Boer-Mulders (BM) and transversity distributions is
more complicated  compared to the Sivers case. The reason
is that the Sivers functions enter the cross section in convolution with
the unpolarized TMD fragmentation functions, \emph{known} from
multiplicities. The BM and transversity functions enter the cross
sections in convolution with the transversely polarized TMD FFs, the
so called Collins functions. The latter can, in principle,  be extracted from
$e^+e^-\to h_1h_2X$, but at present are rather poorly known.

In our analysis we shall consider Collins functions, alongside with
BM and transversity functions, as unknown quantities. In the
difference cross sections  only the valence quark functions
 of both  parton densities and fragmentation functions enter the expressions.

The distribution of transversely polarized quarks  $\qup$ in an unpolarized proton $p$
determines the Boer-Mulders  function \cite{BM}:
\be
 \Delta^N
\! f_{\qup/p}(x_B, \kt) \equiv \Delta f^{q}_{s_y/p}(x_B,\kt)  = - \frac
{\kt}{m_p} \, h_{1}^\perp (x_B, \kt)\,\cdot \label{b-m}
\ee
It is accessed by measuring the BM $\cos 2\phi_h$-asymmetry in unpolarized SIDIS  \cite{general}:
\be
\langle \cos 2\phi_h\rangle^h &\equiv & \frac{ d^4 \langle \cos 2\phi_h\rangle}{d\xb
dQ^2\,dz_h\,dP_T^2}=\frac{\int d\phi_h \,\cos 2\phi_h\,d^5\s^h}{\int
d\phi_h\,d^5\s^h}\nn
& =&\frac{\,\frac{(1-y)}{Q^4}\,F_{UU}^{\cos
2\phi_h}(\xb ,Q^2,z_h,P_T^2)} {\,\frac{1+(1-y)^2}{Q^4}\,F_{UU}(\xb
,Q^2,z_h,P_T^2)},\label{BMasym}
 \ee
  where $d^5 \s^h$ is a short hand
notation for the unpolarized differential SIDIS cross section:
\be
d^5\s^h=\frac{d^5\s_p^h}{dx_B\,dQ^2\,dz_h\,d P_T^2\,d\phi_h}\,\cdot
\ee

The structure function $F_{UU}^{\cos2\phi_h} $ is a convolution  of the
Boer-Mulder (BM) distribution function
  $ \Delta  f^q_{s_y/p}(x,\kt)$ and the  Collins fragmentation function
$ \Delta^N  D_{h/q\uparrow}(z,\pp)$.  The Collins fragmentation
function represents the spin dependent part of the fragmentation
function of a transversely polarized quark with spin polarization
${\bf s}$ and 3-momentum ${\bf p}_q$ into an unpolarized hadron $h$
with transverse momentum ${\bf p}_\perp$ relative to it \cite{C}:
 \be
D_{h/q,s}(z,\pp)=D_{h/q}(z,\pp)+ \frac{1}{2}\, \Delta^N
D_{h/q\uparrow}(z,\pp)\,{\bf \hat s}\cdot({\bf \hat p}_q\times {\bf
\hat p}_\perp)\,\cdot
 \ee
  Often the Collins function is written  as $H_1^\perp (z_h,p_\perp)$ \cite{H_perp}:
\be
\Delta^N
D_{h/q\uparrow}(z,\pp)=\frac{2\pp}{z_h m_h }\,H_1^\perp (z,\pp)
 \ee
 where $m_h$ is the mass of the final hadron.

\section{BM difference asymmetries }

Here we shall work with the difference asymmetries  $\langle\cos
2\phi\rangle^{h-\bar h}$, that we define,
 in analogy to (\ref{BMasym}), through the difference cross sections:
\be
\langle\cos 2\phi\rangle^{h-\bar h}=\frac{\int d\phi_h \,\cos
2\phi_h\,d^5\s^{h-\bar h}}{\int d\phi_h\,d^5\s^{h-\bar h}}
=\frac{\,\frac{(1-y)}{Q^4}\,F_{UU}^{\cos 2\phi_h, h-\bar h}(\xb
,Q^2,z_h,P_T^2)} {\,\frac{1+(1-y)^2}{Q^4}\,F_{UU}^{h-\bar h}(\xb
,Q^2,z_h,P_T^2)}
\ee

  As shown in \cite{we2}, the difference cross sections are determined only by
the valence quark densities. We shall consider the BM asymmetries on  deuteron and proton targets separately.

 \subsection{The standard  parametrization for the deuteron}

 The expressions for the difference asymmetries are especially simple if  we consider SIDIS on a deuteron target. Then

i) independently of the final  hadrons $h-\bar h$, it is always only
the sum of the valence quark BM  distribution that enters:
\be
   \Delta f^{Q_V}_{s_y/p}(\xb ,k_\perp )\equiv \Delta f^{u_V}_{s_y/p}(\xb ,k_\perp )+ \Delta f^{d_V}_{s_y/p}(\xb ,k_\perp )
   \ee

  ii)  for each final hadron $h$, only one combination of valence-quark Collins functions appears

  Thus we need one parametrization for
 the valence BM function $ \Delta f^{Q_V}_{s_y/p}$   and  one for the valence-quark Collins function
 $\Delta^N D_{h/u_V\uparrow}$, or their combination $e_u^2 \Delta^N D_{h/u_V\uparrow} +e_d^2 \Delta^N D_{h/d_V\uparrow}$.

In the currently available  extractions the BM and Collins functions are parametrized proportional to the corresponding
unpolarized transverse-momentum independent collinear  functions, with
factorized flavour independent transverse momentum dependence,
  and flavour-dependent $\xb$  or $z_h$ functions \cite{general,BM_2}.
 In accordance with this we write:
 \be
\hspace*{-.5cm}  \Delta  f^{Q_V}_{s_y/p}(\xb,\kt ,Q^2) \!&=&\! \Delta  f^{Q_V}_{s_y/p}(\xb,Q^2)\;
\sqrt{2e}\,\frac{\kt}{M\BM} \; \frac{e^{-\kt^2/\avk_{BM}
}}{\pi\avk_{BM}}\nn
\hspace*{-.5cm} \Delta  f^{Q_V}_{s_y/p}(\xb,Q^2)\!&=&\! 2\,{\cal
N}\BM^{Q_V}(\xb)\,Q_V(\xb,Q^2) \label{BM-Vdist}
 \ee
 \be
\hspace*{-.5cm} \Delta^N  D_{h/u_V\uparrow}(z_h,\pp ,Q^2) \!&=&\!\Delta^N  D_{h/u_V\uparrow}(z_h,Q^2)\,
\sqrt{2e}\,\frac{\pp}{M\C  } \; \frac{e^{-\pp^2/\avp\C }}{\pi\avp\C
}\,,\quad h=\pi^+,K^+\nn
\hspace*{-.5cm}\Delta^N D_{h/u_V\uparrow}(z_h,Q^2)\!&=&\!2\,{\cal N}^{h/\!u_V}\C
(z_h)\,D_{u_V}^h(z_h,Q^2) \label{Coll-Vfrag}
 \ee
 where $\Delta  f^{Q_v}_{s_y/p}(\xb)$
, $\Delta^N  D_{h/u_V\uparrow}(z_h)$ , ${\cal
N}^{Q_V}\BM(\xb)$ and ${\cal N}^{h/\!u_V}\C (z_h)$)  are unknown functions,
$M\BM$ and $M\C  $ are unknown,   fitted  parameters (often $M\C  $ is denoted by
$M$~\cite{T_1} or $M_h$~\cite{general,T_3}), and
\be
 \avk \BM= \frac{\avk  \, M^2\BM}{\avk  + M^2 \BM}\,
,\qquad \avp\C  =\frac{\avp  \, M\C ^2}{\avp  +M\C ^2}\,\cdot
\label{Coll-frag2}
 \ee

\subsection{The structure functions on deuterons}

For the expressions $F_{UU}^{\cos 2\phi_h,h-\bar h}$  we obtain:
\be
F_{UU}^{\cos 2\phi_h,\pi^+-\pi^-} & = & {\cal K} \BM (z_h,P_T^2)  \,
 \Delta f^{Q_V}_{s_y/p}(\xb ,Q^2)\,[(e_u^2-e_d^2)\,
\Delta^N  D_{\pi^+/u_V\uparrow}(z_h)]\label{FUUBM-pions}\\
F_{UU}^{\cos 2\phi_h,K^+-K^-} & = & {\cal K} \BM (z_h, P_T^2)  \,
 \Delta f^{Q_V}_{s_y/p}(\xb ,Q^2)\,[\,e_u^2\,
\Delta^N  D_{K^+/u_V\uparrow}(z_h)]\label{FUUBM-kaons}\\
F_{UU}^{\cos 2\phi_h,h^+-h^-} & = & {\cal K} \BM (z_h, P_T^2)  \,
 \Delta f^{Q_V}_{s_y/p}(\xb ,Q^2)\nn
 &&\,\times\,[\,e_u^2\,
\Delta^N  D_{h^+/u_V\uparrow}(z_h)+e_d^2\, \Delta^N
D_{h^+/d_V\uparrow}(z_h)]\label{FUU_diff}
\ee
where  ${\cal K}\BM$
is a common flavour-independent factor:
\be
 {\cal
K}\BM(z_h,P_T^2)&=&-e\,A\BM\,A_{Coll}  \,\,  P_T^2 \, \frac{e^{-P_T^2/\avPT
\BM}}{\pi\avPT ^3 \BM}\,z_h\,\cdot
\ee
 Here $A\BM$ and $A_{Coll}  $ are
 flavour and $z_h$-independent factors, determined by the
transverse-momentum dependencies:
 \be
 A\BM
=\frac{\avk^2\BM}{M\BM\avk },\qquad A_{Coll}  =\frac{\avp^2\C}{M\C \avp }.\label{ABM_AC}
 \ee
 and
\be
 \avPT\BM = \avp\C  + z_h^2 \, \avk\BM \,\cdot\label{PTBM}.
\ee
The expressions  in (\ref{FUU_diff}) differ only by the unknown
functions $\Delta^N  D_{h/q_V\uparrow}(z_h)$.

Data on $\langle\cos 2\phi\rangle$ are always presented as function
of one of the kinematic variables
 $(\xb ,z_h,Q^2)$ or $P_T^2$, integrated over the other three.
 We shall next consider  the $\xb$ and $z_h$-dependencies
separately.

\subsection{ The integrated BM asymmetries on deuterons}

Again for simplicity we perform the $P_T$ integrations analytically.
 The general expressions  are presented in the Appendix.
Integrating further we obtain the $\xb$- and $z_h$-dependent BM asymmetries.

 \subsubsection{The $z_h$-dependent BM asymmetries}

 For the $z_h$-dependent BM asymmetries with $h=\pi^+, K^+$ we obtain:
 \be
\hspace*{-.5cm}\langle\cos 2\phi_h\rangle^{h-\bar h}(z_h )
\!=\!{\cal B}\BM^h (z_h)\,\,\frac{z_h}{\avp\C  + z_h^2 \, \avk\BM
}\,\, {\cal N}^{h/u_V}\C   (z_h ) ,\quad h=\pi^+, K^+ \label{BM-z}
\ee
where
\be
\hspace*{-1cm}{\cal B}\BM^h (z_h)
=-2e\,A\BM\,A_{Coll} \,\frac{\int d\xb \,\int
dQ^2\,\frac{1-y}{Q^4}\,\Delta
f_{s_y/p}^{Q_V}(\xb,Q^2)\,D_{u_V}^{h}(z_h,Q^2)} {\int d\xb \,\int
dQ^2\,\frac{1+(1-y)^2}{Q^4}\,Q_V
(\xb,Q^2)\,D_{u_V}^{h}(z_h,Q^2)}\,\cdot
 \ee

As before, for small enough bins in $Q^2$ so that the
$Q^2$-dependence of the collinear parton densities  and FFs can be neglected,
 the FFs cancel and  the formula strongly simplifies:
 \be
\hspace*{-1cm}\langle\cos 2\phi_h\rangle^{h-\bar h}(z_h,\bar Q^2 )
=\bar B\BM(\bar Q^2)\,\,\frac{z_h}{\avp\C + z_h^2 \, \avk\BM }\,
{\cal N}^{h^+/u_V}\C  (z_h ),\quad h =\pi^+, K^+
\label{ABMzh-d}
\ee
 where $\bar{\cal B}\BM$ is independent of
$z_h$ and $h$:
 \be
 \bar B\BM(\bar Q^2) =-e\,A\BM\,A_{Coll} \,\,
\frac{2\,\int d\xb \,\,(1-\bar{y})\,\Delta
f_{s_y/p}^{Q_V}(\xb,\bar Q^2)} {\int d\xb \,\,[1+(1-\bar{y})^2]\,Q_V (\xb,\bar Q^2)}\,\cdot
 \ee
where $\bar Q^2$ is the average $Q^2$-value in the bin and $\bar{y}$ is given in Eq.~(\ref{y}).
The dependence on $h$ in Eq.~(\ref{ABMzh-d}) is only in the $z_h$-dependent part of  Collins function [see Eq.~(\ref{Coll-Vfrag})].

\subsubsection{The $\xb$-dependent BM asymmetries}

For the $\xb$-dependence for each $h=\pi^+, K^+, h^+$ we obtain:
\be
\langle\cos 2\phi_h\rangle^{h-\bar h}(\xb )={\cal
C}\BM ^{h}(\xb)\, {\cal N}^{Q_V}\BM (\xb ),\qquad \forall \,h
\label{cos2phi}
\ee
 where for $h=\pi^+, K^+$ we have:
 \be
  {\cal
C}^{h}\BM (\xb)&=&\!\!-4e\,A\BM\,A_{Coll} \,\frac{\,\int\! \int dQ^2\,
 dz_h\,\frac{1-y}{Q^4} \,Q_V(\xb)\,z_h\,
 \Delta^ND_{h/u_V\uparrow}(z_h)\,/\langle P_T^2\rangle\BM }
{\int \int dQ^2\,dz_h
\,\frac{1+(1-y)^2}{Q^4}\,Q_V(\xb)\,D_{u_V}^{h}(z_h)}\nn
 \ee
  For
unidentified charged  hadrons  $h=h^\pm$ the formula is obtained
with the replacement:
 \be
 \Delta^ND_{h^+/u_V\uparrow}\,\to\,
e_u^2\Delta^ND_{h^+/u_V\uparrow}+e_d^2\Delta^ND_{h^+/d_V\uparrow}\label{replacement_C}
\ee
 and replacement (\ref{replacement}) in the denominator.

 As before, for small enough bins in $Q^2$ this simplifies and one has
\be
\hspace*{-.5cm}\langle\cos 2\phi_h\rangle^{h-\bar
h}(\xb,\bar Q^2 )= \frac{1-\bar{y}}{1 + (1-\bar{y})^2}\,\,\bar C^{h}\BM(\bar
Q^2)\,
 {\cal N}^{Q_V}\BM (\xb ),\quad h=\pi^+, K^+, h^+
\ee
 where $\bar y$ is given in (\ref{y}) and $\bar C\BM^h$ is $\xb$-independent:
\be
\hspace*{-1cm}\bar C^{h}\BM(\bar Q^2)=\!\!-2\,e\,A\BM\,A_{Coll} \,
\frac{ \int dz_h \,z_h\left[\Delta ^N D_{h/u_V\uparrow}(z_h,\bar Q^2)\right]/\langle P_T^2\rangle\BM }
{\int dz_h \,D_{u_V}^{h}(z_h,\bar Q^2)},\quad h=\pi^+, K^+\label{CBMxd}
\ee
\be
\hspace*{-1cm}\bar C^{h^+}\BM(\bar Q^2)=\!\!-2\,e\,A\BM\,A_{Coll} \,
\frac{\int dz_h z_h \!\left[e_u^2\,\Delta^N D_{h^+/u_V\uparrow}(z_h)
+e_d^2\,\Delta^N D_{h^+/u_V\uparrow}(z_h)\right]\!/\langle
P_T^2\rangle\BM } {\int dz_h \left[e_u^2 D^{h^+}_{u_V}+e_d^2
D^{h^+}_{d_V}\right]}\cdot
\ee
In summary, note that

a) as we shall see later,  comparison of $\langle\cos 2\phi_h\rangle^{h-\bar h}(z_h,\bar Q^2),\, h=\pi^+,K^+$
 with the analogous expressions for transversity,
where the same Collins function enterS (see Sect. \ref{Relations 2}),
provides a test of the  factorization scheme approach  currently adopted for
the extraction of the TMD functions. If the test is satisfied, then (\ref{ABMzh-d})
provides a direct measurement for
   ${\cal N}\C^{h/u_V}(z_h)$.

b) $\langle\cos 2\phi\rangle^{h-\bar h}(\xb, \bar
Q^2)$ for $h=\pi^+, K^+,h^+$ provide 3
independent algebraic equations for the $\xb$-behaviour of BM
function ${\cal N}^{Q_V}\BM (\xb )$: the {\it shape} of
$\langle\cos 2\phi\rangle^{h-\bar h}(\xb)$ is
universal for all final hadrons;
 the coefficients  of proportionality  are different for the different final hadrons and
are determined by the valence-quark Collins FFs.

\section{Relations between BM and Sivers asymmetries on deuterons }\label{Relations 1}

In the current analysis an additional simplifying assumption is made, namely
the BM function is assumed proportional  to
its chiral-even partner -- the Sivers function \cite{BM_1,
BM_2}. In our case this reads:
\be
 \Delta
f_{s_y/p}^{Q_V}(x,k_\perp)=\frac{\lambda_{Q_V}}{2}\,\Delta
f_{Q_V/S_T}(x,k_\perp),\label{BM1}
 \ee
 where $\lambda_{Q_V}$ is a fitting parameter.
  This implies that
\be
\avk\BM = \avk\S , \qquad  A\BM = A_{Siv},\qquad {\cal
N}\BM^{Q_V}(\xb)=\frac{\lambda_{Q_V}}{2}\,{\cal
N}^{Q_V}_{Siv}(\xb)\,\cdot\label{BM2}
 \ee

 and
 \be
\avPT\BM = \avPT_{\widetilde{BM}} \equiv \avp\C +z_h^2\avk\S \,\cdot\label{tildeBM}
\ee

Under assumption  (\ref{BM1}), the
 Sivers and BM asymmetries are expressed in terms
of the same valence-quark $\Delta f_{Q_V/S_T}$ Sivers function and differ only by the TMD fragmentation
functions. This  leads to definite relations between
their $\xb$-dependencies, which  we  suggest as possible tests for this assumption.

\subsection{General Relations}

  For the $\xb$-dependence  the  Sivers and BM
asymmetries for  each $h=\pi^+, K^+, h^+$  provide
  independent information about the same $\xb$-dependence
  ${\cal N}_{Siv}^{Q_V}(\xb)$ of the Sivers distribution $\Delta f_{Q_V/S_T}$.
  This leads to the following relation between these asymmetries:
    \be
  {\cal R}_1^{h}(\xb)\equiv
  \frac{\langle\cos\phi_h\rangle^{h-\bar h}(\xb )}{A_{UT}^{Siv,h-\bar h}(\xb)}
  =
  \frac{\lambda_{Q_V}}{2}\,\left[\frac{{\cal C}_{\widetilde{BM}}  (\xb)}{{\cal C}_{Siv}(\xb)}\right]^h,
  \quad h=\pi^+, K^+, h^+
  \ee
 For  $h=\pi^+, K^+$ we have:
  \be
\hspace*{-1cm}\left[\frac{{\cal C}_{\widetilde{BM}} (\xb)}{{\cal
C}_{Siv}(\xb)}\right]^h \!=\!\frac{- 4\sqrt {2e}}{\sqrt \pi}\,
\frac{\int\int
dQ^2\,dz_h\,\frac{1-y}{Q^4}\,Q_V(\xb,Q^2)z_h\,\Delta^N
D_{h/u_V\uparrow}(z_h, Q^2)/\avPT_{\widetilde{BM}} } {\int\int
dQ^2\,dz_h\,\frac{1+(1-y)^2}{Q^4}\,Q_V(\xb,Q^2)z_h\,D_{u_V}^{h}(z_h, Q^2)/\sqrt{\avPT_{Siv}}}\nn
\label{R1test}
\ee
where $\avPT\S$ and $\avPT_{\widetilde{BM}}$ are given in (\ref{PperpSiv}) and (\ref{tildeBM}).
For unidentified charged hadrons $h^\pm$, the result is obtained
with the replacement (\ref{replacement}) and the analogous one for
$\Delta^N D_{h^+/u_V\uparrow}$, Eq.~(\ref{replacement_C}).

The expression for ${\cal R}_1^h$ allows a check of the assumption (\ref{BM1})
between the Sivers and BM functions,  adopted in current  analysis,  without
requiring knowledge of  these functions.
Only the Collins and the collinear valence-quark functions enter Eq.~(\ref{R1test}).

\subsection{Simplification for small $Q^2$-bins}

If the bins in $Q^2$ are small  enough so as to neglect the $Q^2$-evolution of the collinear functions inside a bin, then this ratio
 ${\cal R}_1^{h}$ becomes independent of $\xb$. We have:
 \be
 \label{R1simpletest}
  \bar {\cal R}_1^{h}(\xb,\bar Q^2)=
  \frac{\lambda_{Q_V}}{2}\,\frac{1-\bar{y}}{1 + (1-\bar{y})^2}\,\,\bar C^h(\bar Q^2)
   \ee
  where $\bar y$ is given in (\ref{y}) and $\bar C^h$ is independent of $\xb$:
   \be
   \bar C^h(\bar
Q^2)\equiv \frac{- 8\sqrt {2e}}{\sqrt \pi}\,\frac{\int dz_h\, \,z_h\,    \Delta^N
D_{h/u_V\uparrow}(z_h, \bar{Q}^2     /\avPT_{\widetilde{BM}}} {\int
dz_h\,\,z_h\,\,D_{u_V}^{h}(z_h,\bar Q^2)/\sqrt{\avPT\S}}\,\cdot
 \ee

If data exist for a range of $x_B$ values at the same $\bar{Q}$ then eq.~(\ref{R1simpletest})
  is a remarkably strong requirement and should allow a stringent test of the assumed connection between the Boer-Mulders and Sivers functions.


\section{BM difference asymmetries on protons }

We consider Boer-Mulders difference asymmetries on proton targets only for $\pi^\pm$ and $K^\pm$.
 The expressions are obtained from the  corresponding expressions for deutrons with the following replacements:\\

$\bullet$ for $\underline{\underline{\pi^+-\pi^-}}$

In the numerators, determined by  $F_{UU}^{\cos 2\phi_h,\pi^+-\pi^-}$ (eq. (\ref{FUUBM-pions})):
\be
(e_u^2-e_d^2)\,\Delta f_{s_y/p}^{Q_V}(\xb,Q^2)\, \rightarrow \,\Delta f_{s_y/p}^V(\xb ,Q^2)
\ee
 where
 \be
\Delta f_{s_y/p}^V(\xb ,Q^2)\equiv 2\left[\,e_u^2\,{\cal N}\BM^{u_V}(\xb)\,u_V(\xb,Q^2)-e_d^2\,{\cal N}\BM^{d_V}(\xb)\,d_V(\xb,Q^2)\right]
\ee

$\bullet$ for $\underline{\underline{K^+-K^-}}$

In the numerators, determined by $F_{UU}^{\cos 2\phi_h,K^+-K^-}$ (eq. (\ref{FUUBM-kaons})):
\be
\Delta f_{s_y/p}^{Q_V}(\xb,Q^2)\to \Delta  f_{s_y/p}^{u_V}(\xb,Q^2)
\ee
where
\be
\Delta f_{s_y/p}^{u_V}(\xb,Q^2)=2\,{\cal N}\BM^{u_V}(\xb)\,u_V(\xb,Q^2),
\ee
In the denominators the  replacements for $\pi^+-\pi^-$ and $K^+-K^-$  are (\ref{pions}) and (\ref{K}),  respectively.

Bellow we shall not present all results, but only those  for small enough ranges of $Q^2$,
 where the scale dependence of the parton densities and FFs can be neglected.\\

{\bf 1a)} Again for the $z_h$-dependence we obtain the same behaviour as for deuterons:
\be
\hspace*{-1cm}\langle\cos 2\phi_h\rangle^{h-\bar h}_p(z_h,\bar Q^2 )
=\bar B\BM^h(\bar Q^2)\,\frac{z_h}{\avp \C + z_h^2 \, \avk\BM }\, {\cal N}^{h/u_V}\C (z_h ),
\quad h =\pi^+, K^+\label{ABMzh-p}
\ee
The difference is only in the constant common factors:
\be
\hspace*{-1cm}\bar B\BM^{\pi^+}(\bar Q^2) &=&-e\,A\BM\,A\C\, \frac{2\,\int d\xb \,(1-\bar y)\,\Delta f_{s_y/p}^{V}(\xb,\bar Q^2)}
{\int d\xb \,[1+(1-\bar y)^2]\,f_V (\xb,\bar Q^2)}\nn
\bar B\BM^{K^+}(\bar Q^2) &=&-e\,A\BM\,A\C\, \frac{2\,\int d\xb \,(1-\bar y)\,\Delta f_{s_y/p}^{u_V}(\xb,\bar Q^2)}
{\int d\xb \,[1+(1-\bar y)^2]\,u_V (\xb,\bar Q^2)}\,\cdot
\ee

{\bf 1b)} The $\xb$-dependent BM asymmetries on protons are analogous as those on deuterons,
but with  different combinations of valence-quark densities for pions and kaons.
We have:
\be
\hspace*{-.5cm}\langle\cos 2\phi_h\rangle^{\pi^+-\pi^-}_p(\xb,\bar Q^2 )
&=& \frac{1-\bar{y}}{1 + (1-\bar{y})^2}\,\,\bar C^{\pi}\BM(\bar Q^2)\,
 \frac{\Delta f^V_{s_y/p} (\xb,\bar Q^2)}{2\,f^V(\xb,Q^2)},\nn
 \hspace*{-.5cm}\langle\cos 2\phi_h\rangle^{K^+-K^-}_p(\xb,\bar Q^2 )
&=& \frac{1-\bar{y}}{1 + (1-\bar{y})^2}\,\,\bar C^{K}\BM(\bar Q^2)\,
 {\cal N}^{u_V}\BM (\xb ), \label{ABMxb-p}
\ee
 where $\bar y$ is given in (\ref{y}) and $\bar C\BM^h$ are $\xb$-independent
and  the same as for deuterons, eq.(\ref{CBMxd}),
 $f^V(\xb)$ is known from DIS. These measurements provide  information about
 $\Delta f\BM^V$ and ${\cal N}\BM^{u_V}$.

 In Sect.VI we suggested the ratio of the $\xb$-dependent BM and Sivers
 asymmetries on a deuteron target as a simple test for the used assumption of the proportionality of  the BM and Sivers functions.
   On a proton target, however,
  such a test is not possible for all final hadrons.
  If the BM and Sivers functions are proportional, then only the kaon  BM and Sivers difference asymmetries
   will measure the same ${\cal N}_{Siv}^{u_V}(\xb)$, which
   implies that the ratio of $A_{Siv,p}^{K^+-K^-}(\xb)$
   and $\langle\cos 2\phi_h\rangle_p(\xb )^{K^+-K^-}$ should be independent of $\xb$.

\section{Transversity distributions}

 The distribution of transversely polarized quarks $\qup$ in a transversely polarized proton $\pup$ defines the transversity distributions $h_{1q}(x)$
 or $ \Delta _T q(\xb)$:
\be
 \Delta _T q(\xb) =  h_{1q}(\xb) = \int \!d^2 {\bkt} \,
h_{1q}(\xb,\kt)\label{int-transv} \,.
\ee
 where $h_{1q}(\xb,\kt)$ is the
transversity distribution depending  on the  parton transverse
momentum $\bf k_\perp$, $k_\perp = \vert {\bf k_\perp }\vert$.

The transversity distributions $h_{1q}(x)$ are one of the three
fundamental parton densities, alongside with the unpolarized $q(x)$
and helicity $\Delta q(x)$ distributions, that describe partons in
polarized nucleon. Transversity distributions are the least known
distributions as, being chiral-odd, they cannot be measured in DIS \cite{Barone}.

\section{ Collins difference asymmetries }

At present, knowledge about transversity comes from the single-spin
Collins asymmetries in SIDIS with transversely polarized target
$l+p^\uparrow \to l+h+X$:
 \be
A_{UT}^{\sin(\phi_h+\phi_S),h}&=&\frac{1}{S_T}\, \frac{\int d\phi_h
d\phi_S \left[d^6 \s^\uparrow-d^6\s^\downarrow
\right]^{h}\,\sin(\phi_h+\phi_S)} {\int d\phi_h d\phi_S \left[d^6
\s^\uparrow+d^6\s^\downarrow \right]^{h}}\nn
&=&\frac{\frac{[1-y]}{Q^4}\,F_{UT}^{\sin(\phi_h+\phi_S),h}}{\,\frac{[1+(1-y)^2]}{Q^4}\,F_{UU}^h}\,\cdot
\ee
 The projection using  $\sin(\phi_h+\phi_S)$ singles out the structure function $F_{UT}^{\sin(\phi_h+\phi_S),h}$,
 which depends on a  convolution of the  transversity
 distribution $h_{1q}(x,\kt)$ and the Collins fragmentation function $\Delta^N D_{h/q\uparrow}(z,\pp)$.

The Collins asymmetry for the difference cross section we define
analogously:
\be
A_{UT}^{\sin(\phi_h+\phi_S),{h-\bar
h}}&=&\frac{1}{S_T}\, \frac{\int d\phi_h d\phi_S \left[d^6
\s^\uparrow-d^6\s^\downarrow \right]^{ h-\bar
h}\,\sin(\phi_h+\phi_S)} {\int d\phi_h d\phi_S \left[d^6
\s^\uparrow+d^6\s^\downarrow \right]^{h-\bar h}}\nn
&=&\frac{\frac{[1-y]}{Q^4}\,F_{UT}^{\sin(\phi_h+\phi_S),h-\bar h}}
{\frac{[1+(1-y)^2]}{Q^4}\,F_{UU}^{h-\bar h}}
 \ee
 The expressions for
the structure functions depend on parametrizations for the
transversity and Collins functions, and are different for
different targets. Again we here consider only a deuteron target.


 \subsection{The standard parametrization and  structure functions for deuterons}

In the difference cross sections on a deuteron target only the sum of
the valence-quark transversity distribution   enters:
\be
h_{1Q_V}(\xb,\kt)\equiv h_{1u_V}(\xb,\kt)+h_{1d_V}(\xb,\kt)
\ee
We follow the standard parametrization  presently used in extracting the transversity
from SIDIS data.
 This implies  a standard Gaussian-type transverse-momentum dependence \cite{T_1}:
 \be
h_{1Q_V}(\xb,\kt)=h_{1Q_V}(\xb)\,\frac{e^{-\kt^2/\avk\T}}{\pi\avk\T},
\ee
where  $\avk\T$ is a new parameter and the following parametrization is used for the distribution
$h_{1Q_V}(\xb)$  \cite{T_2}:
 \be
h_{1Q_V}(\xb)=\frac{1}{2}\,{\cal
N}\T^{Q_V}(\xb)\,\left[Q_V(\xb)+\Delta Q_V(\xb)\right]
 \ee
where, as earlier,  $Q_V(\xb)=u_V+d_V$ and $\Delta Q_V(\xb)= \Delta u_V+\Delta d_V$  are  the sum of  the collinear
unpolarized and longitudinally polarized valence-quark parton  densities, while  ${\cal N}\T^{Q_V}(\xb)$ is the
new unknown function.

The parametrization for the Collins function was given in
(\ref{Coll-Vfrag}).
 With these currently used standard parametrizations  we obtain:
\be
F_{UT}^{\sin (\phi_h +\phi_S),\pi^+-\pi^-} & = & {\cal K} \T
(z_h,P_T^2)  \,
  h_{1Q_V}(\xb,Q^2)\,[(e_u^2-e_d^2)\,\Delta^N D_{\pi^+/u_V\uparrow}(z_h,Q^2)
  \label{FUT-T-pions}\\
F_{UT}^{\sin (\phi_h +\phi_S),K^+-K^-} & = & {\cal K} \T (z_h,
P_T^2)  \,
  h_{1Q_V}(\xb,Q^2)\,[\,e_u^2\,\Delta^N D_{K^+/u_V\uparrow}(z_h,Q^2)
  \label{FUT-T-kaons}\\
F_{UT}^{\sin (\phi_h +\phi_S),h^+-h^-} & = & {\cal K} \T (z_h,
P_T^2)  \,
 h_{1Q_V}(\xb,Q^2)\nn
 &&\times\,[\,e_u^2\,\Delta^N D_{\pi^+/u_V\uparrow}(z_h,Q^2)
 +e_d^2\,\Delta^N D_{\pi^+/u_V\uparrow}(z_h,Q^2)]\nn
\ee
 where
 \be
  {\cal K}\T(z_h,P_T^2)&=&\frac{\sqrt{2e}}{2}\,A_{Coll} \,
P_T \, \frac{e^{-P_T^2/\avPT \T}}{\pi\avPT ^2 \T} ,
\nn
\langle P_T^2\rangle\T&=&\avp\C +z_h^2\avk\T\,\cdot\label{PTT}
 \ee
 Here $\avp\C $ was given in Eq.~(\ref{Coll-frag2}) and $A_{Coll} $ in (\ref{ABM_AC}).
  The unknown quantities are $\avk\T$ and $\avp\C$ in
${\cal K} \T (z_h,P_T^2)$, and $h_{1Q_V}(\xb)$. They are the same for
all final hadrons.

In the next
Sections we discuss the $\xb$- and $z_h$-dependent Collins asymmetries.

\subsection{The integrated  Collins asymmetries on deuterons}

 The integration over $P_T$ we perform analytically. The expressions are given in the Appendix.
  \\

\subsubsection{The $z_h$-dependent Collins asymmetries}

For the $z_h$-dependence of Collins asymmetry with $h=\pi^+, K^+$ we obtain:
\be
 A_{UT}^{Coll,h-\bar h}\!(z_h)&=&\,\frac{{\cal
B}^h\T(z_h)}{\sqrt{\avp\C +z_h^2\avk\T }} \,{\cal N}\C ^{h/u_V}(z_h
),\quad h=\pi^+, K^+ \label{C-z}
\ee
where
\be
{\cal B}^h\T(z_h)=\sqrt{\frac{e\pi}{2}}\, A_{Coll}  \,\frac{\int
\int\,d\xb\,dQ^2\,\frac{1-y}{Q^4} \,
h_{1Q_V}(\xb,Q^2)\,D_{u_V}^{h}(z_h,Q^2)} {\int\int
d\xb\,dQ^2\,\frac{1+(1-y)^2}{Q^4}\,Q_V(\xb,Q^2)\,D_{u_V}^{h}(z_h,Q^2)}
\ee

As before, for reasonably small bins in $Q^2$, the expressions for the  $z_h$-dependence of the
asymmetries considerably simplify:
\be
A_{UT}^{Coll,h-\bar h}\!(z_h,\bar Q^2)=\,\frac{\bar B\T(\bar
Q^2)}{\sqrt{\avp\C +z_h^2\avk\T }} \,{\cal N}\C ^{h/u_V}(z_h ),\quad
h=\pi^+,K^+\label{ACollzh-d}
 \ee
where $\bar B\T$ is independent of both $z_h$ and $h$:
\be
 \bar  B\T(\bar Q^2)=\sqrt{\frac{e\pi}{2}}\, A_{Coll}  \,\frac{\int
\,d\xb\, \, (1-\bar{y}) \,h_{1Q_V}(\xb,\bar Q^2)}
{\int d\xb\,[1+(1-\bar{y})^2]\,Q_V(\xb,\bar Q^2)}
 \ee

Thus, the $z_h$-behaviour of the Collins asymmetry provides independent
measurements for
   ${\cal N}\C^{h/u_V}(z_h)$, for $h=\pi^+,K^+$. The dependence on
 the type of the final hadron is only in ${\cal N}\C^{h/u_V}$.

\subsubsection{The $\xb$-dependent Collins asymmetries}

 For the $\xb$-dependent Collins asymmetry for any
$h=\pi^+, K^+, h^+ $ we have:
\be
A_{UT}^{Coll,h-\bar h}(\xb)={\cal C}\T^{h}(\xb)\, {\cal N}\T^{Q_V}(\xb),\quad h=\pi^+, K^+, h^+.
\ee
For $h=\pi^+,K^+$ the functions ${\cal C}\T^{h}(\xb)$  are:
\be
\hspace*{-1cm}{\cal
C}\T^{h}(\xb)=\frac{1}{4}\,\sqrt{\frac{e\pi}{2}}\,A_{Coll}  \,\frac{\int
dQ^2\,\frac{1-y}{Q^4}\,(Q_V+\Delta Q_V)(\xb)\, \int dz_h
\left[\Delta^N D_{h/ u_V\uparrow}(z_h)\right]/\sqrt{\langle
P_T^2\rangle\T} } {\int
dQ^2\frac{1+(1-y)^2}{Q^4}\,Q_V(\xb,Q^2)\,\int dz_h
\,D_{u_V}^{h}(z_h)}\,\cdot
\ee
For unidentified
charged  hadrons   ${\cal C}\T^{h^+}(\xb)$  is obtained with the
replacements (\ref{replacement_C}) and (\ref{replacement}).

Again, for small enough bins in  $Q^2$ we obtain
 \be
 A_{UT}^{Coll,h-\bar h}(\xb,\bar
Q^2)=\bar C\T^{h}(\bar Q^2)\,\frac{1-\bar{y}}{1 + (1-\bar{y})^2}\left[1 + \frac{\Delta Q_V(\xb,\bar Q^2)}{Q_V(\xb,\bar Q^2)}\right]\,
{\cal N}\T^{Q_V}(\xb)\,.
 \ee

The dependence on the final hadrons is only  in the constant factor
$\bar C^h\T$:
\be
\hspace*{-1cm}
\bar C^h\T(\bar Q^2)&=&\frac{1}{4}\,\sqrt{\frac{e\pi}{2}}\,A_{Coll}  \,\, \frac{\int dz_h
\left[\Delta^N D_{h/ u_V\uparrow}(z_h,\bar
Q^2)\right]/\sqrt{\langle P_T^2\rangle\T} }
{\int dz_h \,D_{u_V}^{h}(z_h,\bar Q^2)},\quad h=\pi^+,K^+\label{barC-T}\\
\hspace*{-1cm}
\bar C^{h^+}\T\!(\bar Q^2)&=&\frac{1}{4}\,\sqrt{\frac{e\pi}{2}}\,A_{Coll}  \,\, \frac{\int dz_h\,
\left[e_u^2\Delta^N D_{h^+/ u_V\uparrow}+e_u^2\Delta^N D_{h^+/
u_V\uparrow}\right]/\sqrt{\langle P_T^2\rangle\T} } {\int dz_h
\left[e_u^2 D^{h^+}_{u_V}+e_d^2 D^{h^+}_{d_V}\right]},
 \ee
where $\langle P_T^2\rangle\T$ is given in Eq.~(\ref{PTT}).

In general, the asymmetries $A_{UT}^{Coll,h-\bar h}\!(z_h)$ and
$A_{UT}^{Coll,{ h-\bar h}}\!(\xb)$ for each final hadron  provide 2
independent measurements  for the 2 unknown quantities, the BM and
Collins functions, without any assumptions about the sea-quark TMD
densities and fragmentation functions; also
 only the best known valence-quark
collinear densities enter the expressions.

\section{Relations between BM and Collins asymmetries on deuterons}\label{Relations 2}

Common for BM and transversity distributions is that they enter SIDIS cross section convoluted to
the same Collins fragmentation functions. Using the standard parametrizations, this implies that in BM and transversity
difference asymmetries only one Collins function enters, the same in both asymmetries -- $\Delta D^N_{\pi^+/u_V\uparrow}$ and
 $\Delta D^N_{K^+/u_V\uparrow}$ for $\pi^+-\pi^-$ and $K^+-K^-$ respectively, and it factorizes. They differ only by
 the different
 TMD parton densities, namely the transversity or BM distributions.
 This holds both for deuteron and proton targets and  allows to obtain
 direct connections between the $z_h$-behaviour of the BM and transversity asymmetries, which function as
  direct tests of the adopted factorized parametrizations.
  Here we present the results for deuteron target.  The results for proton target and the relations between proton and deuteron
  targets will
  be given in Section \ref{proton_Coll}.

\subsection{General Relations}

The $z_h$-dependencies of BM and Collins
asymmetries, Eqs.~ (\ref{BM-z}) and (\ref{C-z}),  for each
$h=\pi^+, K^+$
 provide 2 completely different measurements for the same
 Collins function. Eliminating the Collins function,  yields the following general relation between the asymmetries:
  \be
  {\cal R}_2^h(z_h)&\equiv&\frac{\langle\cos 2\phi_h\rangle^{h-\bar h}(z_h )}{A_{UT}^{Coll,h-\bar h}(z_h)}\nn
  &=&
  \frac{z_h\,\sqrt{\avPT\T}}{\avPT\BM}\,\left[\frac{{\cal B}\BM(z_h)}{{\cal B}\T(z_h)}\right]^h
  ,\quad h^\pm=\pi^\pm,K^\pm\label{BMColl}
  \ee
   where
  \be
\left[\frac{{\cal B\BM}(z_h)}{{\cal
B}\T(z_h)}\right]^h=\frac{-2\sqrt{2e}}{\pi}\,A\BM\,
\frac{\int\int\,d\xb\,dQ^2\,\frac{1-y}{Q^4}\,\Delta
f_{s_y/p}^{Q_V}(\xb,Q^2)\,D_{u_V}^{h^+}(z_h,Q^2)}
{\int\int\,d\xb\,dQ^2\,\frac{1-y}{Q^4}\,h_{1,Q_V}(\xb,Q^2)\,D_{u_V}^{h^+}(z_h,Q^2)}
\ee

\subsection{Simplification for small $Q^2$-bins}

 For small enough bins in $Q^2$,
the $z_h$-dependence
 is completely fixed and is independent of the final hadrons for $h=\pi^+, K^+$:
\be
 {\cal R}_2^h(z_h,\bar Q^2)=
  \frac{z_h\,\sqrt{\avPT\T}}{\avPT\BM}\,\frac{\bar B\BM (\bar Q^2)}{\bar B\T(\bar Q^2)}
  ,\quad h=\pi^+, K^+
\label{BMColl1}
  \ee
  where $\bar B\BM/\bar B\T$ is  independent of  both $z_h$ and $h$:
  \be
\frac{\bar B\BM}{\bar B\T}=\frac{-2\sqrt{2e}}{\pi}\,A\BM\,
\frac{\int \,d\xb\,(1-\bar{y})\,\Delta
f_{s_y/p}^{Q_V}(\xb,\bar Q^2)}
{\int\,d\xb\,(1-\bar{y})\,h_{1,Q_V}(\xb,\bar
Q^2)}\,\cdot
\ee
The expressions for $\avPT\T$ and $\avPT\BM$ are given in (\ref{PTT}) and (\ref{PTBM}). Any dependence on $h$ would
be an indication that the $p_\perp$ dependence of the Collins function, as controlled by the value of
$\avp\C $ is not hadron-type independent.

These relations provide a  general  test of the
standardly used parametrizations. If these tests are fulfilled , we can use them to obtain or test information
about  the transversity and BM distributions without knowledge of the
Collins functions.

\subsection{Implications of the  commonly used assumptions}

 As the presently available data are not adequate to fully determine  the BM and transversity distributions,
 additional assumptions are used.  For the BM
function this is relation (\ref{BM1}) between the BM and Sivers
functions, and eqs. (\ref{BM2}) and (\ref{tildeBM}) that follow.
 For the transversity distribution the usual assumption is \cite{T_1, T_3,T_2}:
\be
 \avk\T=\avk  , \label{kT}
 \ee
 which implies
\be
\avPT\T = \avPT_{\widetilde T} \equiv \avp\C +z_h^2\avk .
\ee

With these assumptions,  relation (\ref{BMColl}) reads:
 \be
 {\cal R}_2^h(z_h)=
  \frac{\lambda_{Q_V}}{2}\,
  \frac{z_h\,\sqrt{\avPT_{\widetilde T}}}
  {\avPT_{\widetilde{BM}}}\,\left[\frac{{\cal B}_{\widetilde{BM}}(z_h)}{{\cal B}\T(z_h)}\right]^h
  ,\quad h=\pi^+,K^+
  \ee
    where
  \be
\left[\frac{{\cal B}_{\widetilde{BM}}(z_h)}{{\cal
B}\T(z_h)}\right]^h=\frac{-2\sqrt{2e}}{\pi}\,A_{Siv}\,
\frac{\int\int\,d\xb\,dQ^2\,\frac{1-y}{Q^4}\,\Delta
f_{Q_V/S\T}(\xb,Q^2)\,D_{u_V}^{h^+}(z_h,Q^2)}
{\int\int\,d\xb\,dQ^2\,\frac{1-y}{Q^4}\,h_{1,Q_V}(\xb,Q^2)\,D_{u_V}^{h^+}(z_h,Q^2)}
\ee
and $\avPT_{\widetilde{BM}}$ was given in (\ref{tildeBM}).

For small enough bins in $Q^2$ relation (\ref{BMColl1}) becomes independent of the final
hadron,
 with a completely fixed $z_h$-dependence:
  \be
 {\cal R}_2^h(z_h,\bar Q^2)=
  \frac{\lambda_{Q_V}}{2}\,
  \frac{z_h\,\sqrt{\avPT_{\widetilde T}}}{\avPT_{\widetilde{BM}}}\,
  \left[\frac{\bar B_{\widetilde{BM}}(\bar Q^2)}{\bar B\T (\bar Q^2)}\right]
  ,\quad h=\pi^+,K^+
  \ee
where $\bar{\cal B}_{\widetilde{BM}}/\bar{\cal B}\T$ is constant in
$z_h$, independent of the final hadron:
\be
\left[\frac{\bar
B_{\widetilde{BM}}}{\bar B\T}\right]=\frac{-2\sqrt{2e}}{\pi}\,A_{Siv}\,
\frac{\int\,d\xb\,\,(1-\bar{y})\,\Delta
f_{Q_V/S\T}(\xb,\bar Q^2)}
{\int\,d\xb\,(1-\bar{y})\,h_{1,Q_V}(\xb,\bar
Q^2)}\,\cdot
\ee


\section{The Collins difference asymmetries on protons }\label{proton_Coll}

We consider Boer-Mulders difference asymmetries for $\pi^\pm$ and $K^\pm$.
 The expressions are obtained from the  corresponding expressions for deuterons with the following replacements :\\

$\bullet$ for $\underline{\underline{\pi^+-\pi^-}}$

In the numerators, determined by eq.  (\ref{FUT-T-pions}):
\be
(e_u^2-e_d^2)\,h_{1Q_V}(\xb,Q^2)\, \rightarrow \,h_{1V}(\xb ,Q^2)
\ee
 where
 \be
h_{1V}(\xb ,Q^2)&=& \left[\,e_u^2\,h_{1u_V}(\xb,Q^2)-e_d^2\,h_{1d_V}(\xb,Q^2)\right]\nn
h_{1u_V}&=&\frac{1}{2}\,{\cal N}_T^{u_V}(\xb)\,[u_V+\Delta u_V]\nn
h_{1d_V}&=&\frac{1}{2}\,{\cal N}_T^{d_V}(\xb)\,[d_V+\Delta d_V]
\ee

$\bullet$ for $\underline{\underline{K^+-K^-}}$

In the numerators, determined by eq.  (\ref{FUT-T-kaons}):
\be
h_{1Q_V}(\xb,Q^2)\, \rightarrow \,h_{1u_V}(\xb ,Q^2)
\ee
In the denominators the  replacements for $\pi^+-\pi^-$ and $K^+-K^-$  are (\ref{pions}) and (\ref{K}),  respectively.

Again, we shall not present all results but only those  for small enough ranges of $Q^2$,
 where the $Q^2$-evolution of the parton densities and FFs can be neglected.
 We shall discuss in more details only the $z_h$-dependent asymmetries, as they provide interesting tests for the assumed Gaussian,
 flavour and hadron independent, transverse-momentum dependence.
  \\

 {\bf 1a)}  As before, for reasonably small bins in $Q^2$, the expressions for the  $z_h$-dependence of the
asymmetries considerably simplify. They are similar to those of the deuteron:
\be
A_{UT,p}^{Coll,h-\bar h}\!(z_h,\bar Q^2)=\,\frac{\bar B^h_{T,p}(\bar
Q^2)}{\sqrt{\avp\C +z_h^2\avk\T }} \,{\cal N}\C ^{h/u_V}(z_h ),\quad
h=\pi^+,K^+\label{ACollzh-p}
 \ee
the only difference is that  the $\bar B^h_{T,p}$ depend on $h$:
\be
\bar  B^{\pi^+}_{T,p}(\bar Q^2)&=&\sqrt{\frac{e\pi}{2}}\, A_{Coll}  \,\frac{\int
\,d\xb\, \, (1-\bar{y}) \,h_{1V}(\xb,\bar Q^2)}
{\int d\xb\,[1+(1-\bar{y})^2]\,f_V(\xb,\bar Q^2)}\nn
\bar  B^{K^+}_{T,p}(\bar Q^2)&=&\sqrt{\frac{e\pi}{2}}\, A_{Coll}  \,\frac{\int
\,d\xb\, \, (1-\bar{y}) \,h_{1u_V}(\xb,\bar Q^2)}
{\int d\xb\,[1+(1-\bar{y})^2]\,u_V(\xb,\bar Q^2)}
 \ee

Common to  the expressions for the $z_h$-dependent BM and Collins asymmetries on proton and deuteron targets,
eqs. (\ref{ABMzh-d}), (\ref{ABMzh-p}),  (\ref{ACollzh-d}) and (\ref{ACollzh-p}),  is that the Collins function enters
all expressions. This allows us to obtain
useful relations between them. Below, for simplicity, we shall use the notation:
\be
A^{BM,h-\bar h}\equiv \langle\cos 2\phi_h\rangle^{h-\bar h}\,\cdot
\ee

i) The ratio of the $z_h$-asymmetries on protons and deuterons are independent of $z_h$:
\be
\frac{A_d^{Coll,h-\bar h}(z_h)}{A_p^{Coll,h-\bar h}(z_h)}=\left(\frac{B\T}{B^h_{T,p}}\right),\qquad
\frac{A_d^{BM,h-\bar h}(z_h)}{A_p^{BM,h-\bar h}(z_h)}=\left(\frac{B\BM}{B^h_{BM,p}}\right),\quad h=\pi^+, K^+
\ee

ii) The ratios of the $z_h$-dependent BM and Collins asymmetries on protons and deuterons
 have the same, completely fixed
$z_h$-behaviour, determined solely by the Gaussian parameters $\avp\C, \avk\T$ and $\avk\BM$:
\be
\frac{A_d^{BM,h-\bar h}(z_h)}{A_d^{Coll,h-\bar h}(z_h)}\propto \frac{A_p^{BM,h-\bar h}(z_h)}{A_p^{Coll,h-\bar h}(z_h)}
\propto \frac{A_p^{BM,h-\bar h}(z_h)}{A_d^{Coll,h-\bar h}(z_h)}\propto \frac{A_d^{BM,h-\bar h}(z_h)}{A_p^{Coll,h-\bar h}(z_h)}
\propto  \frac{z_h\,\sqrt{\avPT\T}}{\avPT\BM},\quad  h=\pi^+, K^+
\ee
These are clear predictions that could test the assumed Gaussian-type dependence on the transverse-momentum.
The expressions for $\avPT\T$ and $\avPT\BM$ are given in (\ref{PTT}) and (\ref{PTBM}). The dependence on
$h$  is only in the constant common factors.  For example:
\be
  \frac{A_p^{BM,h-\bar h}(z_h)}{A_p^{Coll,h-\bar h}(z_h)}=
  \frac{z_h\,\sqrt{\avPT\T}}{\avPT\BM}\,\left(\frac{\bar B_{BM,p} (\bar Q^2)}{\bar B_{T,p}(\bar Q^2)}\right)^h
  ,\quad h=\pi^+, K^+
\label{BMColl1}
  \ee
  where $(\bar B_{BM,p}/\bar B_{T,p})^h$ is independent of  $z_h$, but different for the different final hadrons $h$:
  \be
\left(\frac{\bar B_{BM,p}}{\bar B_{T,p}}\right)^{\pi^+}&=&\frac{-2\sqrt{2e}}{\pi}\,A\BM\,
\frac{\int \,d\xb\,(1-\bar{y})\,
\Delta f_{s_y/p}^{V}(\xb,\bar Q^2)}
{\int\,d\xb\,(1-\bar{y})\,h_{1V}(\xb,\bar
Q^2)}\,\nn
\left(\frac{\bar B_{BM,p}}{\bar B_{T,p}}\right)^{K^+}&=&\frac{-2\sqrt{2e}}{\pi}\,A\BM\,
\frac{\int \,d\xb\,(1-\bar{y})\,
\Delta f_{s_y/p}^{u_V}(\xb,\bar Q^2)}
{\int\,d\xb\,(1-\bar{y})\,h_{1u_V}(\xb,\bar
Q^2)}\,\cdot
\ee
\\

{\bf 1b)}  The $\xb$-dependent Collins asymmetries are obtained straight forward with  the above replacements.
For the  asymmetries with pions and kaons we obtain:
\be
A_{UT,p}^{Coll, \pi^+-\pi^-}(\xb ,\bar Q^2)&=&\bar C\T^{\pi}\,\frac{1-\bar y}{1+(1-\bar y)^2}\,
\frac{e_u^2\,{\cal N}\T^{u_V}(\xb)(u_V+\Delta u_V)- e_d^2\,{\cal N}\T^{d_V}(\xb)(d_V+\Delta d_V)}
{e_u^2u_V-e_d^2d_V}\\
A_{UT,p}^{Coll, K^+-K^-}(\xb ,\bar Q^2)&=&\bar C\T^{K}\,\frac{1-\bar
y}{1+(1-\bar y)^2}\, \left(1+\frac{\Delta u_V}{u_V}\right)\,{\cal
N}\T^{u_V}(\xb)
\ee
 where $\bar C^h\T$ are the same as for deuteron, given by (\ref{barC-T}).
.
\section{Conclusions}

We have studied the  Sivers, Boer-Mulders and transversity asymmetries for the
difference cross sections of hadrons with opposite charges, in SIDIS reactions.
  We have considered  deuteron and proton targets separately
$l+d \to l+h^\pm +X$ and $l+p \to l+h^\pm +X$.

There are 3 essential properties of the difference cross sections  that
 follow from the   of charge and isospin invariance of the strong interactions
  and are crucial in obtaining our results: \newline
i)  for the parton densities  there is no sum over quark
flavours, but only one combination of the valence-quark TMD-distributions enters.
\newline
ii) for the fragmentation functions only the
valence-quark TMD fragmentation function for each type of final
hadron appears \newline
  iii) there is no contribution from  sea quarks

At present, data on these asymmetries are presented for the
integrated asymmetries i.e.the $\xb$- and $z_h$-dependent
asymmetries. In calculating  the asymmetries we have followed the simplified approach currently used
   in extracting the various functions from data i.e. we work in LO in QCD and use
   the standard factorized parametrizations for the TMD functions. As we have demonstrated,  these simplifications can be tested. If the tests are succesful we have  shown that  the difference asymmetries on  deuterons
determine  the valence-quark $Q_V=(u_V+d_V)$ Sivers, BM and
transversity  distributions,  and the difference asymmetries on protons determine the Sivers, BM and
transversity  distributions for the combinations
$e_u^2\Delta f_{u_V}-e_d^2 \Delta f_{d_V}$ for pions, and $\Delta f_{u_V}$ for kaons.
 Measurements on both proton and deuteron targets will allow to determine the valence-quark $u_V$ and $d_V$ TMD densities
 separately and free from the contributions of the  sea-quarks densities  and fragmentation functions.

 The remarkably simple and powerful tests of the whole approach,  based on the standard factorized parametrizations,  depend on having data  available in reasonably small bins in $Q^2$,  such  that the $Q^2$-evolution  of the relevant
 collinear parton densities and fragmentation functions  can be neglected inside each  bin, which leads to dramatic simplifications:

1)   The $z_h$-behaviour of $A_{Siv}^{h-\bar h} (z_h)$,
  both for proton and deuteron targets, is a totally fixed function of $z_h$, the same for all final hadrons.
   It is determined solely by the Gaussian-type $k_\perp$-dependence; the  FFs completely cancel.

2) The $\xb$-behaviour of $A_{Siv}^{h-\bar h} (\xb)$  on deuterons, for
 different hadrons, is  proportional to the same
 Sivers ${\cal N}_{Siv}^{Q_V}(\xb )$ function.
This  implies that the ratio for  different hadrons should be independent of $\xb$:
 \be
\frac{A_{Siv}^{\pi^+-\pi^-} (\xb,\bar Q^2)}{A_{Siv}^{K^+-K^-} (\xb,\bar Q^2)}=\frac{\bar C_{Siv}^{\pi^+}(\bar Q^2)}{\bar C_{Siv}^{K^+}(\bar Q^2)}
\ee
Analogously for the BM asymmetry $\langle\cos 2\phi\rangle^{h-\bar h} (\xb)$:
 \be
\frac{\langle\cos 2\phi\rangle^{\pi^+-\pi^-} (\xb,\bar Q^2)}{\langle\cos 2\phi\rangle^{K^+-K^-} (\xb,\bar Q^2)}=\frac{\bar C\BM^{\pi^+}(\bar Q^2)}{\bar C\BM^{K^+}(\bar Q^2)}
\ee
and for the Collins asymmetry $A\T^{h-\bar h} (\xb)$:
 \be
\frac{A\T^{\pi^+-\pi^-} (\xb,\bar Q^2)}{A\T^{K^+-K^-} (\xb,\bar Q^2)}=\frac{\bar C\T^{\pi^+}(\bar Q^2)}{\bar C\T^{K^+}(\bar Q^2)},
\ee
with  analogous relations   when $\pi^\pm$ or $K^\pm$ are replaced by $h^\pm$.

3)  The ratio of the $z_h$-dependent BM and Collins asymmetries  on protons and deuterons
 is a totally fixed function of $z_h$,
the same for $h=\pi^+,K^+$.  It is determined solely by the Gaussian-type $k_\perp$-dependence;
the  Collins FFs completely cancel.

4) A common assumption in the present analysis is that the  BM
functions are  proportional to the Sivers functions. As a test of
this relation we suggest the ratio of the $\xb$-dependent BM and
Sivers asymmetries  on deuterons. If the  assumed relationship is
valid, this ratio  should be independent of $\xb$.
  On a proton target, this assumption can be tested only for the  kaon asymmetries.

Failure to satisfy these tests implies that the extraction of the various distributions and fragmentation functions cannot be trusted.
  If a test is passed successfully then use of the
   measured difference asymmetries  allows the extraction of the Sivers, BM and transversity
   \emph{valence-quark} distributions with better precision, free from the uncertainties of the sea or  strange quark contributions.\newline

\section*{Acknowledgements }  E.C. is grateful to Oleg Teryaev for helpful discussions,
to Bakur Parsamyan for stressing the experimental importance of the data on
proton target, and for the hospitality of  JINR  where this work was
finalized.
 E.L is grateful to The Leverhulme Trust for an Emeritus Fellowship.

 \section*{Appendix }

Here we give the analytic expressions for the $P_T$-integrated difference asymmetries on deuteron.
 We perform the $P_T$-integration using the formula:
 \be
 \int_0^\infty
d\,P_T^2\,\,\pt\,\frac{e^{-P_T^2/\avPT_S}}{\pi\avPT_S^2}=\frac{1}{2\sqrt{\pi}
\sqrt{\avPT_S}}
\ee
 \be
\int_0^\infty d\,P_T^2\,\,\frac{e^{-P_T^2/\avPT}}{\pi\avPT} =\frac{1}{\pi}
\ee
 \be
 \int_0^\infty
d\,P_T^2\,\,\pt^2\,\frac{e^{-P_T^2/\avPT_{BM}}}{\pi\avPT_{BM}^3}=\frac{1}{\pi
\avPT_{BM}}\,\cdot\label{PT2}
\ee

For the general expression of  Sivers asymmetry $A_{UT}^{Siv,{
\pi^+-\pi^-}}$, after integrating over $P_T^2$ we obtain:
\be
\hspace*{-1.2cm}A_{UT}^{Siv,{
\pi^+-\pi^-}}\!&=&\frac{\sqrt{e\pi}}{4\sqrt{2}}\, A_{Siv}\,
\frac{\frac{A(y)}{Q^4}\,\Delta f_{Q_V/S_T}(\xb,Q^2)\,z_h \left[
D_{u_V}^{\pi^+}(z_h,Q^2)\right]/\sqrt{\langle P_T^2\rangle\S }}
{\frac{A(y)}{Q^4}\,Q_V(\xb ,Q^2)\,D_{u_V}^{\pi^+}(z_h,Q^2)}\\
\hspace*{-1.2cm}A_{UT}^{Siv,{
K^+-K^-}}\!&=&\frac{\sqrt{e\pi}}{4\sqrt{2}}\, A_{Siv}\,
\frac{\frac{A(y)}{Q^4}\,\Delta f_{Q_V/S_T}(\xb,Q^2) \,z_h \left[
D_{u_V}^{K^+}(z_h,Q^2)\right]/\sqrt{\langle P_T^2\rangle\S }}
{\frac{A(y)}{Q^4}\,Q_V(\xb ,Q^2)\,D_{u_V}^{K^+}(z_h,Q^2)}\\
\hspace*{-1.2cm}A_{UT}^{Siv,{
h^+-h^-}}\!&=&\frac{\sqrt{e\pi}}{4\sqrt{2}}\, A_{Siv}\,
\frac{\frac{A(y)}{Q^4}\, \Delta f_{Q_V/S_T}(\xb) \,z_h \left[
e_u^2\,
 D^{h^+}_{u_V}(z_h)+e_d^2\,  D^{h^+}_{d_V}(z_h)\right]/\sqrt{\langle P_T^2\rangle\S }}
{\frac{A(y)}{Q^4}\,Q_V(\xb)\,[\,e_u^2\,
 D^{h^+}_{u_V}(z_h)+e_d^2\,  D^{h^+}_{d_V}(z_h)]}\nn
\ee

 The BM asymmetries , integrated over $P_T^2$ are:
\be
\hspace*{-1cm} \langle\cos 2\phi\rangle^{\pi^+ -
\pi^-}&=&-e\,A\BM\,A_{Coll} \,\frac{\frac{(1-y)}{Q^4}\,
  \Delta f^{Q_V}_{s_y/p}(\xb)\,[z_h\,\Delta^N  D_{\pi^+/u_V\uparrow}(z_h)]/\avPT_{BM}}
{\frac{1+(1-y)^2}{Q^4}\,Q_V(\xb )\,D_{u_V}^{\pi^+}(z_h)}\label{1}\\
\hspace*{-1cm}\langle\cos
2\phi\rangle^{K^+-K^-}&=&-e\,A\BM\,A_{Coll} \,\frac{\frac{(1-y)}{Q^4}\,
 \Delta f^{Q_V}_{s_y/p}(\xb)\,[z_h\,\Delta^N  D_{K^+/u_V\uparrow}(z_h)/\avPT_{BM}]}
{\frac{1+(1-y)^2}{Q^4}\,Q_V(\xb )\,D_{u_V}^{K^+}(z_h)}\label{2}\\
\hspace*{-1cm}\langle\cos
2\phi\rangle^{h^+-h^-}&=&-e\,A\BM\,A_{Coll} \,\nn
&&\hspace*{-2cm}\times\frac{\frac{(1-y)}{Q^4}\,
 \Delta f^{Q_V}_{s_y/p}(\xb)\,z_h[\,e_u^2\,
\Delta^N  D_{h^+/u_V\uparrow}(z_h)+e_d^2\, \Delta^N
D_{h^+/d_V\uparrow}(z_h)]/\avPT_{BM}}
{\frac{1+(1-y)^2}{Q^4}\,Q_V(\xb )\left[\,e_u^2\,D_{u_V}^{h^+}(z_h)+
e_d^2\,D_{d_V}^{h^+}(z_h)\right]}\label{3}
\ee

The Collins asymmetries after integration over $P_T^2$ are:
\be
\hspace*{-1.2cm}A_{UT}^{Coll,{
\pi^+-\pi^-}}\!&=&\frac{\sqrt{e\pi}}{2\sqrt 2}\, A_{Coll} \,
\frac{\frac{[1-y]}{Q^4}\,h_{1Q_V}(\xb,Q^2 )\,\Delta^N
D_{\pi^+/u_V\uparrow} (z_h,Q^2)\,/\sqrt{\langle P_T^2\rangle\T }}
{\frac{[1+(1-y)^2]}{Q^4}\,Q_V(\xb,Q^2)\,D_{u_V}^{\pi^+}(z_h,Q^2)}\\
\hspace*{-1.2cm}A_{UT}^{Coll,{
K^+-K^-}}\!&=&\frac{\sqrt{e\pi}}{2\sqrt 2}\, A_{Coll} \,
\frac{\frac{[1-y]}{Q^4}\,h_{1Q_V}(\xb,Q^2 )\,\Delta^N
D_{K^+/u_V\uparrow} (z_h,Q^2)\,/\sqrt{\langle P_T^2\rangle\T }}
{\frac{[1+(1-y)^2]}{Q^4}\,Q_V(\xb,Q^2)\,D_{u_V}^{K^+}(z_h,Q^2)}\\
\hspace*{-1.2cm}A_{UT}^{Coll,{
h^+-h^-}}\!&=&\frac{\sqrt{e\pi}}{2\sqrt 2}\, A_{Coll} \,
\frac{\frac{[1-y]}{Q^4}\,h_{1Q_V}(\xb ) \left[ e_u^2\Delta^N
D_{h^+/u_V\uparrow}(z_h)+e_d^2\,\Delta^N
D_{h^+/d_V\uparrow}(z_h)\right]/\sqrt{\langle P_T^2\rangle\T }}
{\frac{[1+(1-y)^2]}{Q^4}\,Q_V(\xb )\,[\,e_u^2\,
 D^{h^+}_{u_V}(z_h)+e_d^2\,  D^{h^+}_{d_V}(z_h)]}\nn
\ee



\begin{thebibliography}{99}

\bibitem{Bacchetta}  A. Signori, A.Bacchetta, M.Radici and G.Schnell, JHEP {\bf 11} (2013) 194

\bibitem{Q2-evol}  S. Mert Aybat, Ted C. Rogers, Phys.Rev. {\bf D83} (2011) 114042 (arXiv:1101.5057);
 S. Mert Aybat, John C. Collins, Jian-Wei Qiu, Ted C. Rogers, Phys. Rev. {\bf D 85} (2012) 034043 (arXiv:1110.6428);
  M. Anselmino, M. Boglione, S. Melis, Phys. Rev. {\bf D 86} (2012) 014028 (ArXiv:1204.1239);
  Umberto D'Alesio, Miguel G. Echevarria, Stefano Melis, Ignazio Scimemi, JHEP11(2014)098 (arXiv:1407.3311)

\bibitem{we1} E. Christova and E.Leader, Nucl. Phys. {\bf B607} (2001) 369 (arXiv:hep-ph/0007303);
\\
 Phys. Rev.  D  {\bf 79} (2009) 014019 ( arXiv:0809.0191)

 \bibitem{we2} E. Christova, Phys. Rev.    {\bf D 90}  (2014)  054005  (arXiv:1407.5872);

 \bibitem{HERMES_2009}  A.Airapetian et al, Phys.Rev.Lett. {\bf 103} (2009) 152002

\bibitem{Sivers} D.Sivers, Phys. Rev. {\bf D41} (1990) 83; {\bf 43}(1991) 261

\bibitem{MuldersTangerman} P.J.Mulders and R.D.Tangerman, Nucl.Phys. {\bf B461} (1996) 197
 (arXiv:hep-ph/9510301)

\bibitem{Anselmino_06} M. Anselmino et al., Phys. Rev.  D {\bf 73} (2006) 014020 (hep-ph/0509035)

\bibitem{general} M. Anselmino,  M. Boglione, U. D'Alesio, S. Melis, F. Murgia, E.R. Nocera and A. Prokudin,
 Phys. Rev.  D {\bf 83} (2011) 114019 (arXiv:1101.1011)


 \bibitem{Anselmino_08} M. Anselmino et al., Eur.Phys.J. {\bf A39} (2009) 89
(arXiv:0805.2677)

\bibitem{multpls} M. Anselmino,  M. Boglione,, J.O.Gonzalez H., S.Melis and A.Prokudin,
 arXiv:1312.6261

 \bibitem{BM_2} Vincenzo Barone, Stefano Melis and Alexej Prokudin,
  Phys. Rev. {\bf D 81}, 114026 (2010)

 \bibitem{BM} D. Boer and P.J.Mulders, Phys. Rev.  {\bf D 57} 5780 (1998)

 \bibitem{C} J.C.Collins, Nucl. Phys. {\bf B 396} 161 (1993)

\bibitem{H_perp} D. Boer, R.Jacob and P.Mulders,
Nucl. Phys. {\bf B 504} 345 (1997); \\ A.Bacchetta et al., JHEP 0702;093 (2007)

 \bibitem{T_1} M. Anselmino et al., Phys. Rev.  {\bf D 75} (2007) 054032 (arXiv:hep-ph/0701006)

 \bibitem{T_3} M. Anselmino et al., Phys. Rev.  {\bf D 87} (2013) 094019 (arXiv:1303.3822)

\bibitem{BM_1} Vincenzo Barone, Alexei Prokudin and Bo-Qiang Ma, Phys.
Rev. {\bf D 78}, 045022 (2008)

\bibitem{Barone} V.Barone, A.Drago and P.G.Ratcliffe, Phys. Rep. {\bf 359}, 1 (2002)

\bibitem{T_2} M. Anselmino et al.,  Nucl.Phys.Proc.Suppl. {\bf 191}, 98 (2009) (arXiv:0812.4366)


 \end{thebibliography}
 \end{document}